\documentclass[%
 aip,
 jmp,%
 amsmath,amssymb,
 reprint,%
]{revtex4-1}

\usepackage{graphicx}
\usepackage{dcolumn}
\usepackage{bm}
\usepackage{amsmath}
\usepackage{color}
\usepackage{amssymb}
\usepackage{wasysym}
\usepackage{amsthm}
\usepackage{soul}
\usepackage{mathtools}   

\usepackage{float}
\usepackage[caption = false]{subfig}

\newcommand*{\bfrac}[2]{\genfrac{}{}{0pt}{}{#1}{#2}}

\begin{document}

\date{\today}


\title[]{Cholesterics of colloidal helices: Predicting the macroscopic pitch from the particle shape and thermodynamic state}

\author{Simone Dussi}
\email{s.dussi@uu.nl}
\affiliation{Soft Condensed Matter, Debye Institute for Nanomaterials Science, Utrecht University, Princetonplein 5, 3584 CC Utrecht, The Netherlands}

\author{Simone Belli}
\author{Ren\'e van Roij}
\affiliation{Institute for Theoretical Physics, Center for Extreme Matter and Emergent Phenomena, Utrecht University, Leuvenlaan 4, 3584 CE Utrecht, The Netherlands}

\author{Marjolein Dijkstra}
\email{m.dijkstra1@uu.nl}
\affiliation{Soft Condensed Matter, Debye Institute for Nanomaterials Science, Utrecht University, Princetonplein 5, 3584 CC Utrecht, The Netherlands}

\begin{abstract}
Building a general theoretical framework to describe the microscopic origin of macroscopic chirality in (colloidal) liquid crystals is a long-standing challenge. Here, we combine classical density functional theory with Monte Carlo calculations of virial-type coefficients, to obtain the equilibrium cholesteric pitch as a function of thermodynamic state and microscopic details. Applying the theory to hard helices, we observe both right- and left-handed cholesteric phases that depend on a subtle combination of particle geometry and system density. In particular, we find that entropy alone can even lead to a (double) inversion in the cholesteric sense of twist upon changing the packing fraction. We show how the competition between single-particle properties (shape) and thermodynamics (local alignment) dictates the macroscopic chiral behavior. Moreover, by expanding our free-energy functional we are able to assess, quantitatively, Straley's theory of weak chirality, used in several earlier studies. Furthermore, by extending our theory to different lyotropic and thermotropic liquid-crystal models, we analyze the effect of an additional soft interaction on the chiral behavior of the helices. Finally, we provide some guidelines for the description of more complex chiral phases, like twist-bend nematics. Our results provide new insights on the role of entropy in the microscopic origin of this state of matter.

\end{abstract}



\maketitle

\section{Introduction}
\begin{figure}[h!t]
\begin{center}
\includegraphics[width=0.5\textwidth]{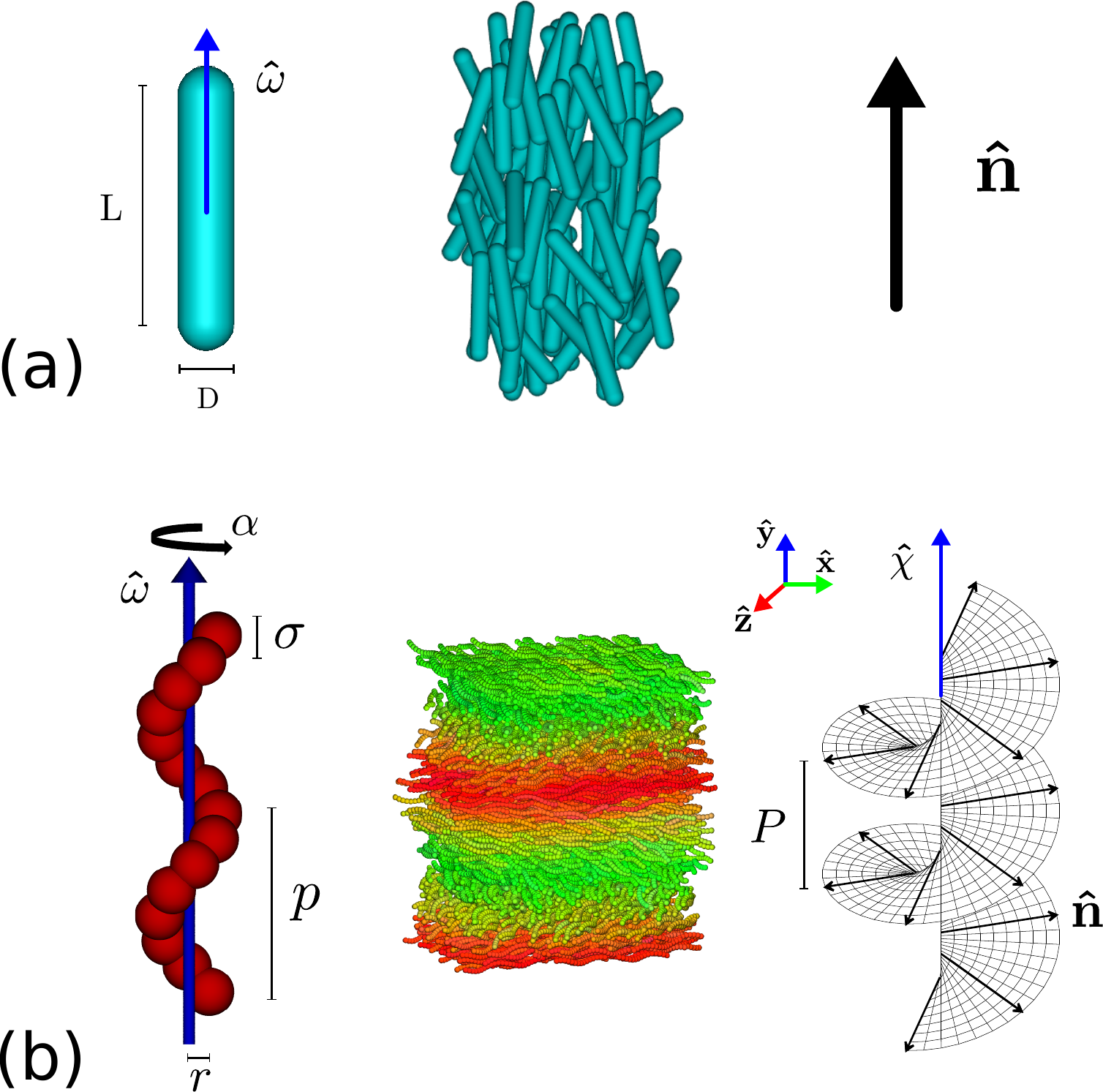}
\end{center}
\caption{Microscopic particle model (left), macroscopic liquid crystalline phase (cartoon in the middle), and schematic of the nematic director field (right). {\bf (a)} Achiral colloids are often modeled as spherocylinders of length $L$ and diameter $D$, whose orientation is described by a unit vector $\mathbf{\hat{\omega}}$. These rod-like particles give rise to a uniaxial and achiral nematic phase with a uniform nematic director $\mathbf{\hat{n}}$. {\bf (b)} The hard helix model consists of $N_s$ hard spheres of diameter $\sigma$, rigidly fused together to form an helix with molecular pitch $p$ and radius $r$. Particle orientation is described by a unit vector $\mathbf{\hat{\omega}}$ and an internal angle $\alpha$. Colloidal helices self-assemble into a cholesteric configuration. The nematic director field $\mathbf{\hat{n}}$ exhibits an helical arrangement along the chiral director $\mathbf{\hat{\chi}}$, with periodicity given by the cholesteric pitch $P$.}
\label{fig:micro_macro}
\end{figure}

``What is the origin of chirality in the cholesteric phase of virus suspensions?''.~\cite{grelet2003} With such an intriguing question, which is to date unanswered, Eric Grelet and Seth Fraden titled their paper about a decade ago. The link between micro- and macro-chirality remains elusive not only in virus suspensions but in many systems exhibiting liquid crystal phases.~\cite{degennes} These thermodynamic states in between the disordered liquid phase and the (fully) ordered crystal phase, consist of anisotropic particles or molecules featuring long-range orientational order but no (or only partial) positional order. In this study, we focus on nematic phases, where the particles are preferentially aligned along a common direction, identified by a unit vector called nematic director~$\mathbf{\hat{n}}$, while keeping their centers of mass homogeneously distributed in space. Whereas the nematic director of an ordinary (achiral and uniaxial) nematic phase is homogeneously distributed throughout the system (Fig.~\ref{fig:micro_macro}{\bf(a)}), the cholesteric phase, often called chiral nematic, displays an helical arrangement of the director field (Fig.~\ref{fig:micro_macro}{\bf(b)}). The typical length scale associated to this macroscopic chirality, that determines the periodicity of such an imaginary helix, is named cholesteric pitch~$P$. Depending on the twist sense of the director field around the chiral director $\mathbf{\hat{\chi}}$, the liquid crystal phase is denoted right- or left-handed.

Cholesterics are readily observed in both thermotropic molecular compounds and lyotropic colloidal suspensions.~\cite{degennes} The former class of liquid crystals, in which phase transitions are mainly governed by temperature, has found wide technological application in the opto-electronic industry due to the unique combination of rheological, electrical and optical properties conferred by the chiral structure.~\cite{collings,tamaoki,lagerwall} Derivatives of cholesterol,~\cite{dupre,huff,rossi} the first liquid-crystal-forming systems experimentally observed,~\cite{reinitzer,lehmann} belong to this class.

Several lyotropic systems, where the phase behavior is density-driven,~\cite{vroege} such as suspensions of colloidal particles or polymers, exhibit chiral order as well. Examples range from biological materials, such as DNA,~\cite{robinson,livolant,zanchetta} filamentous viruses,~\cite{dogic2000,grelet2003,dogic2006,barry2009,zhangz,tombolato2006} cellulose and derivatives,~\cite{revol,werbowyj} chiral micelles;~\cite{hiltrop} to synthetic polymers, such as polyisocyanates~\cite{aharoni,sato} and polysilanes.~\cite{watanabe} Suspensions of filamentous viruses are among the most studied~\cite{dogic2006} colloidal systems in which chirality plays a major role in self-assembly processes at different levels, leading to fascinating phenomena.~\cite{gibaud,sharma} At the microscopic level, charged (protein) subunits self-assemble into supramolecular helical structures. However, whereas suspensions of fd virus particles exhibit chirality also at macroscopic scale, thereby stabilizing a cholesteric phase,~\cite{grelet2003,dogic2000,dogic2006,barry2009} other virus particles, such as tobacco mosaic virus~\cite{bawden} and Pf1 virus~\cite{dogic2000} particles, with similar helical charge distributions, form only a uniaxial nematic phase, challenging the idea that molecular chirality is a guarantee for macroscopic chirality. Even though in the latter case the cholesteric pitch is expected to be too large to be directly observed in experiments, very subtle differences at the single-particle level can drastically change the macroscopic self-organization.~\cite{barry2009,zhangz} Surprisingly, fd virus particles that are sterically stabilized with a polymer coat so thick that the electrostatic chiral interactions are proved to be fully masked, still exhibit a cholesteric phase.~\cite{grelet2003} These observations, together with a recent study aimed to map the fd-virus phase diagram onto that of hard rods,~\cite{grelet2014} suggest that entropy alone could govern the phase behavior of (some) virus suspensions, including the stabilization of a cholesteric phase. However, whereas the phase diagram of the coated fd virus and the nematic order parameter were independent of ionic strength, the cholesteric pitch varied surprisingly strongly with ionic strength.~\cite{grelet2003} 

The underlying competition between steric and electrostatic interactions appears even more evident in suspensions of another charged filamentous viruses (M13), where a left-handed cholesteric phase was obtained by right-handed particles.~\cite{tombolato2006} The observed sense of the macroscopic twist could not be correctly predicted by modeling the particles as hard bodies without taking into account a soft electrostatic contribution.~\cite{tombolato2006} Cholesterics with opposite handedness with respect to that of the constituent particles were also observed in solutions of ultrashort DNA.~\cite{tombolato2005,zanchetta} Surprisingly, a peculiar type of DNA oligomers showed an inversion in the helical sense of the cholesteric phase upon changing system concentration, suggesting that packing arguments could explain in which sense these systems should twist. However, the cholesteric pitch seemed to be influenced by other factors as well, such as particle length and oligomer sequences, but not, for example, by particle flexibility. As a result, no simple rules could exhaustively explain the chiral behavior.~\cite{zanchetta} Inversion in the cholesteric handedness has been reported in other studies as well, often concerning temperature-driven systems.~\cite{huff,katsonis} The list of systems indicating that the connection between micro- and macro-chirality is far from trivial, is quite long,~\cite{dalmas} and difficult to rationalize due to the different interactions in place.

On the other hand, a recent study seemed to have succeeded in identifying a chiral system ruled by entropic effects only.~\cite{barry2006} Indeed, interactions in suspensions of helical flagella extracted from bacteria can be finely tuned by modifying solvent properties. In particular conditions, these colloidal particles can be ultimately considered as hard helices, whose exact shape can be also precisely regulated.~\cite{barry2006,hasegawa} As expected, when helical flagella self-assemble, the chirality is transmitted to the liquid crystalline state. However, the formed chiral nematic phase has a different symmetry from the cholesteric phase and was identified as a conical phase.~\cite{barry2006,meyer,kamien} Why such a phase should be thermodynamically more stable than the cholesteric is another question thickening the mystery of colloidal chirality.  An even more complicated mechanism of chirality propagation from molecular to macroscopic scale has been observed in thermotropic bent-core liquid crystals.~\cite{eremin} In peculiar cases the intricate coupling between twisting and bending deformations stabilize another chiral nematic phase, named twist-bend nematic.~\cite{borshch} In this instance the local nematic director $\mathbf{\hat{n}}$ is tilted with respect to the chiral director $\mathbf{\hat{\chi}}$, resembling therefore the conical order.

In view of such a complex and sometimes controversial experimental scenario, it is not surprising that a unifying microscopic theory is still lacking. The attempt of incorporating all the interactions present in experiments in a suitable model for chiral particles, is often beyond the limits of current theoretical tools and computational resources. For this reason, despite very few exceptions,~\cite{tombolato2005} focus was separately given to the chiral behavior arising from either purely hard-core repulsions~\cite{straley,evansgt,varga2011,wensink2014} or from soft electrostatic potentials only.~\cite{varga2006,wensink2009,wensink2011,wensink2014} Even with this simplification, the complexity of a chiral interparticle potential is such that most of these studies resorted to coarse-grained potentials,~\cite{germano,varga2006,memmer2001,wensink2009,wensink2011,wensink2014} in which the microscopic chiral features are masked into a single pseudo-scalar parameter.~\cite{goossens} By contrast, we decided to build our study in small steps,~\cite{belli} first improving our understanding of entropy-driven systems, and only at a later stage introducing further elements into the particle model. 

Beside the particle model, an appropriate theoretical approach is crucial. Computer simulation methods are limited by the large number of particles required to accommodate a full rotation of the nematic director. Despite ad-hoc techniques that have been developed to try to overcome such an issue,~\cite{allen1993mol,frenkel} only few simulation studies, mainly using coarse-grained potentials tailored to minimize cholesteric pitches, have been dedicated to the investigation of cholesteric phases.~\cite{memmer2001,berardi,varga2003,varga2006,germano,allen1993pre,camp} Therefore, to shed new light on the microscopic origin of the macroscopic chirality, we appeal to a suitable microscopic theory. A successful example of a microscopic theory, often used in soft matter, is due to Onsager, who was the first in 1949 to explain the role of entropy on the liquid-crystalline behavior of anisotropic particles.~\cite{onsager} However, it took until 1976 to describe the cholesteric ordering, when Straley proposed his approach, based on theory of elastic deformations.~\cite{straley} The seminal work of Straley has been extensively used in modern studies to predict the cholesteric pitch of several colloidal systems.~\cite{emelyanenko,wensink2009,wensink2011,wensink2014,tombolato2005,frezza2014,kornyshev2002,odijk} Only few exceptions presented alternative methods, limited, however, by severe analytical assumptions.~\cite{evansgt,varga2011} Despite the undoubted relevance of Straley's pioneering work, his approach is based on two main assumptions. First, the theory is rigorously valid only in the limit of weak macroscopic chirality, a limit that is anyway usually not far from the experimental conditions. Second, the theory cannot be solved fully self-consistently, in the sense that the orientation distribution of the cholesteric phase equals that of the underlying uniaxial achiral nematic phase, thereby neglecting the differences in the local order between the latter and a cholesteric phase. Of course this second assumption is consistent with the perturbative treatment of chirality in Straley's theory. Additionally, Straley's approach has been used only for the description of cholesterics but not extended to the study of chiral nematics with different symmetries (e.g. twist-bend or conical phases). 

We have recently introduced a novel approach to address these issues.~\cite{belli} Our aim is twofold: by refreshing the theoretical description of chiral nematics within the density functional framework, we propose an additional tool to advance our understanding of this complex state of matter. At the same time, by applying our theory to hard helices, we provide new insights into the role of entropy in colloidal cholesterics. Indeed, such an apparently simple model exhibits a fairly rich and complex chiral behavior,~\cite{belli,frezza2014} that goes beyond simplified scenarios suggested in earlier studies.~\cite{straley,harris1997,harris1999} Moreover, despite a thorough simulation study aimed to map out the entire phase diagram,~\cite{frezza2013,kolli2014jcp,kolli2014soft} leading to a newly observed chiral nematic (screw-like) phase, a question mark is still pending on the cholesteric phase and a definitive evidence from simulations is yet to come.

This paper is organized as follows: we dedicate Sec.~\ref{sec:theory} to readers interested in technical details, where we describe extensively the theoretical framework used and its numerical implementation. In Sec.~\ref{sec:results}, we study the cholesteric order in systems of hard helices. In Sec.~\ref{sec:soft}, we analyze the effect of an additional soft, short-range interaction on the macroscopic chiral behavior, thereby providing an explicit example of how the theory can be applied to different particle models. Furthermore, in Sec.~\ref{sec:twistbend}, we discuss some guidelines to extend the theory to (more complex) chiral nematic phases of different symmetry than the cholesteric one, focusing in particular on the study of twist-bend nematics. We conclude our paper with some final remarks in Sec.~\ref{sec:conclusions}.

\section{Theory}
\label{sec:theory}
\subsection{Revisiting Onsager theory: uniaxial colloids and nematic order}
\label{sec:onsager}
To introduce our formalism, we first revisit Onsager's theory~\cite{onsager} within the framework of classical density functional theory (DFT),~\cite{evansbob} in the simple case of anisotropic rod-like colloids of diameter $D$ and length $L$ (see Fig.~\ref{fig:micro_macro}{\bf(a)}).

The center-of-mass position of a particle with cylindrical symmetry can be described by a three-dimensional vector $\mathbf{r}=(x,y,z) \in V $, with $V$ the volume of the system, whereas a unit vector $\mathbf{\hat{\omega}}=(\sin\theta \cos\phi, \sin\theta \sin\phi,\cos\theta)$ represents the particle orientation, where $\theta \in [0,\pi)$ and $\phi \in [0,2\pi)$ are the polar and the azimuthal angle with respect to $\mathbf{\hat{n}}$. The single-particle density $\rho (\mathbf{r},\mathbf{\hat{\omega}})$ of a generic phase depends on the single-particle degrees of freedom and satisfies the normalization condition
\begin{equation}
\int_V \, d\mathbf{r} \, \oint \, d\mathbf{\hat{\omega}} \, \rho (\mathbf{r},\mathbf{\hat{\omega}}) = N,
\end{equation}
where $N$ is the total number of particles and the rotation-invariant measure is $d\mathbf{\hat{\omega}} = d\phi  \, d\cos\theta$. The free energy is a functional of the single-particle density and can be written as a sum of ideal and excess contributions,
$ \mathcal{F} [\rho(\mathbf{r},\mathbf{\hat{\omega}})] = \mathcal{F}_{id} [\rho] + \mathcal{F}_{ex} [\rho] $. The ideal term reads
\begin{equation}
\beta \mathcal{F}_{id}[\rho]=\int_V \, d\mathbf{r} \, \oint \, d\mathbf{\hat{\omega}} \, \rho (\mathbf{r},\mathbf{\hat{\omega}}) \, [ \log \mathcal{V} \rho (\mathbf{r},\mathbf{\hat{\omega}}) - 1 \, ],
\end{equation}
where $\beta=1/k_BT$, with $k_B$ the Boltzmann constant and $T$ the temperature, and $\mathcal{V}$ is an (irrelevant) constant thermal volume. For the excess part we consider the second-order truncation of the virial expansion (second-virial approximation):
\begin{multline}
\beta \mathcal{F}_{ex}[\rho]= \\-\frac{1}{2} \int_V d\mathbf{r} \, d\mathbf{r'} \oint  d\mathbf{\hat{\omega}} \, d\mathbf{\hat{\omega}'} f(\mathbf{r}-\mathbf{r'},\mathbf{\hat{\omega}},\mathbf{\hat{\omega}'}) \rho (\mathbf{r},\mathbf{\hat{\omega}}) \rho (\mathbf{r'},\mathbf{\hat{\omega}'}),
\end{multline}
where the interactions between particles are contained in the Mayer function
\begin{equation}
f(\mathbf{r}-\mathbf{r'},\mathbf{\hat{\omega}},\mathbf{\hat{\omega}'})= e^{ -\beta U(\mathbf{r}-\mathbf{r'},\mathbf{\hat{\omega}},\mathbf{\hat{\omega}'})} -1,
\end{equation}
where $U(\mathbf{r}-\mathbf{r'},\mathbf{\hat{\omega}},\mathbf{\hat{\omega}'})$ is the pair potential.

In a nematic phase, the positions of the particles are homogeneously distributed throughout the system and the single-particle density can be rewritten as $\rho(\mathbf{r},\mathbf{\hat{\omega}})=n \psi(\mathbf{\hat{\omega}})$, where $n=N/V$ is the average number density and $\psi(\mathbf{\hat{\omega}})$ is the orientation distribution function (ODF). We choose a Cartesian reference frame in which the $z$-axis is parallel to the nematic director $\mathbf{\hat{n}}$. 
Since the nematic director is the symmetry axis for global rotations, the ODF is independent of the azimuthal angle $\phi$ and depends only on the polar angle: $\psi(\mathbf{\hat{\omega}})=\psi(\theta)$. Inserting this into $\mathcal{F}$ and integrating out the spatial and azimuthal degrees of freedom we obtain
\begin{multline}
\label{eq:onsager}
\frac{\beta \mathcal{F}[\psi]}{V}= n (\log \mathcal{V}n -1)+ 2\pi n \int_{-1}^{1} \, d\cos\theta \, \psi(\theta) \, \log \psi(\theta) \, \\
+ \frac{n^2}{2} \int \,  d\cos\theta \, d\cos\theta' \,E(\theta,\theta') \, \psi(\theta) \, \psi(\theta'),
\end{multline}
where we identify the three terms associated to translational entropy, orientational entropy and excess contributions related to the excluded volume 
\begin{equation}
E (\theta,\theta') = - \int \, d\phi \, d\phi' \, d\mathbf{r} \, f (\mathbf{r},\mathbf{\hat{\omega}},\mathbf{\hat{\omega}}) .
\end{equation}
Eq.~(\ref{eq:onsager}) is an exact expression for the free energy of the nematic phase of infinitely long (aspect ratio $L/D \rightarrow \infty$) hard rods, as derived by Onsager.~\cite{onsager} Subsequently, Parsons and Lee~\cite{parsons,lee} used the same approach to describe nematics of rods with finite $L/D$, mapping the system free energy to that of hard spheres at the same packing fraction $\eta=n v_0$, with $v_0=\frac{\pi}{4}LD^2 + \frac{\pi}{6} D^3$ being the single-particle volume. This correction introduces a density-dependent prefactor 
\begin{equation}
G(\eta)=\frac{(1-\frac{3}{4}\eta)} {( 1- \eta)^2}
\end{equation}
 in front of $\mathcal{F}_{ex}$. Following the DFT recipe,~\cite{evansbob} once the free-energy functional is defined, the next step consists of minimizing $\mathcal{F}[\psi]$ with respect to $\psi(\theta)$, subject to the normalization condition $\int d\mathbf{\hat{\omega}} \psi(\mathbf{\hat{\omega}})=1$. In the case of (Parsons-Lee-)Onsager theory, the resulting non-linear equation for $\psi(\theta)$ reads:
\begin{equation}
\label{eq:ons_nnlin}
\psi(\theta)=\frac{1}{Z} \exp \left( - n \, G(\eta) \, \int_{-1}^{1} \, d\cos\theta' \, \frac{E(\theta,\theta')}{2\pi} \, \psi(\theta')   \right), 
\end{equation}
where $Z$ is a normalization constant such that $2\pi \int_{0}^{\pi} d\theta \sin \theta \psi(\theta)=1$. Eq.~(\ref{eq:ons_nnlin}) can be solved \emph{self-consistently} at fixed $n$ and $E(\theta,\theta')$, and the resulting (equilibrium) ODF can be used to calculate all relevant thermodynamic quantities. An example on how to solve it numerically by using a discrete grid for the polar angle $\theta$ in the case of hard rods, can be found in Ref.~[69]
. Theoretical predictions obtained for spherocylinder systems agree well with simulation results.~\cite{bolhuis}

\subsection{Density functional theory for chiral nematics}
\label{sec:dft} 
It is known that a chiral particle cannot be uniaxial.~\cite{harris1997,harris1999} Biaxiality introduces an additional degree of freedom: the orientation of a generic rigid body is described by three scalar parameters, $(\theta,\phi,\alpha) \in [0,\pi) \times [0,2\pi) \times [0,2\pi)$, known as Euler angles, or alternatively by a  $3 \times 3$ rotation matrix $\mathcal{R}$. 
The rotation matrix $\mathcal{R}$ can be parametrized in terms of the unit vector $\mathbf{\hat{\omega}}$, representing the orientation of the main long axis, and the internal azimuthal angle $\alpha$ (cf. Fig.~\ref{fig:micro_macro}{\bf(b)}). The single-particle density has to be modified for the extra degree of freedom and is now subjected to the following normalization condition
\begin{equation}
\int_V \, d\mathbf{r} \, \oint \, d\mathcal{R} \, \rho (\mathbf{r},\mathcal{R}) = N ,
\end{equation}
with $ d\mathcal{R} = d\alpha \,  d\phi \, d\cos\theta$. Since the achiral nematic phase is a homogeneous phase with orientational order only, the corresponding single-particle density depends on the orientation variable only: $\rho(\mathbf{r},\mathcal{R})=n \psi(\mathcal{R})$, with $\int d\mathcal{R} \psi(\mathcal{R})=1$. In the most general case $\psi(\mathcal{R})$ describes a biaxial phase. 

Let us now consider a chiral nematic phase of pitch $P$ with the chiral director $\mathbf{\hat{\chi}}$ aligned along the $y$ axis (cf. Fig~\ref{fig:micro_macro}{\bf(b)}). The cholesteric pitch $P$ is related to the chiral wave vector $q$ through $P=2\pi/q$. The chiral structure implies that the ODF at arbitrary $y$ can be deduced from that at $y=0$ by rotating by an angle $2\pi y /P = qy$ around the $y$-axis. Such a condition reads
\begin{equation}
\label{eq:chiral_distrib}
\rho(\mathbf{r},\mathcal{R})=n \psi \mathbf{(}\, \mathcal{T}_q (\mathbf{r}) \, \mathcal{R} \, \mathbf{)},
\end{equation}
where 
\begin{equation}
\mathcal{T}_q(\mathbf{r}) \equiv \mathcal{R}_{\mathbf{\hat{\chi}}}(q \, \mathbf{\hat{\chi}} \cdot \mathbf{r}) = \mathcal{R}_{\mathbf{\hat{y}}}(qy)
\end{equation}
 is a rotation around the chiral director $\mathbf{\hat{\chi}}$ (that coincides with the $y$-axis) by an angle $qy$. From Eq.~(\ref{eq:chiral_distrib}) we can immediately verify that chiral nematic phases are characterized by the inequality $\rho(\mathbf{r},\mathcal{R}) \neq \rho(-\mathbf{r},\mathcal{R})$.
By explicitly imposing the parity symmetry $\rho(\mathbf{r},\mathcal{R}) = \rho(-\mathbf{r},-\mathcal{R}) $, based on the assumption that the physics does not change by passing from a right-handed to a left-handed reference frame and vice versa, we can rewrite the previous as $\rho(\mathbf{r},\mathcal{R}) \neq \rho(\mathbf{r},-\mathcal{R})$. A necessary (not sufficient) condition for a system of particles to manifest chiral nematic ordering is that an inversion transformation, i.e., $\mathcal{R} \rightarrow -\mathcal{R}$, \emph{does not} transform a particle into itself, as previously noticed.~\cite{harris1997,harris1999,straley} It is also interesting to ask what kind of two-body interaction $U(\mathbf{r},\mathcal{R},\mathcal{R}')$ generates chiral nematic ordering. Again, a necessary condition for chirality is $U(-\mathbf{r},\mathcal{R},\mathcal{R}') \neq U(\mathbf{r},\mathcal{R},\mathcal{R}')$ or, alternatively, $U(\mathbf{r},-\mathcal{R},-\mathcal{R}') \neq U(\mathbf{r},\mathcal{R},\mathcal{R}')$.

Eq.~(\ref{eq:chiral_distrib}) describes the functional dependence of the single-particle density of a chiral nematic phase. The corresponding density functional theory, already briefly outlined in our previous work,~\cite{belli} is based on three steps that will be described in detail here. First of all, we will insert the functional dependence of the chiral phase $\rho(\mathbf{r},\mathcal{R})=\rho(\mathcal{T}_q(\mathbf{r})\mathcal{R})$ into $\mathcal{F}[\rho(\mathbf{r},\mathcal{R})]$. Secondly, we will rewrite $\mathcal{F}[\rho(\mathcal{T}_q(\mathbf{r})\mathcal{R})]$ in such a way that the dependence on $q$ and $\rho(\mathcal{R})$ are disentangled. In other words, we will construct a functional $\mathcal{F}_q$ such that $\mathcal{F}_q [\rho(\mathcal{R})]=\mathcal{F}[\rho(\mathcal{T}_q(\mathbf{r})\mathcal{R})]$. Finally, we will minimize the density functional with respect to $\psi(\mathcal{R})$ and $q$. Let $\rho^*(\mathcal{R})$ be the solution at given $q$ and given number density $n$, such that $F_q(n)=\mathcal{F}_q[\rho^*(\mathcal{R})]$. The equilibrium value of $q$ (and hence the equilibrium cholesteric pitch $P$) at the number density $n$ corresponds to the minimum in $q$ of the free energy $F_q(n)$.

As a lemma, let us first see that the ideal part of the free-energy functional
\begin{multline}
\beta \mathcal{F}_{id}[\rho(\mathcal{T}_q(\mathbf{r})\mathcal{R})]= \\ \int_V \, d\mathbf{r} \, \oint \, d\mathcal{R} \, \rho(\mathcal{T}_q(\mathbf{r})\mathcal{R}) \, [ \log \mathcal{V} \rho(\mathcal{T}_q(\mathbf{r})\mathcal{R})) - 1 \, ] 
\end{multline}
does not contribute to the chiral ordering. By changing the orientation integration variable from $\mathcal{R}$ to $\mathcal{Q}=\mathcal{T}_q(\mathbf{r})\mathcal{R}$ (with unit Jacobian), we obtain
\begin{multline}
\label{eq:chiralfid}
\beta \mathcal{F}_{id}[\rho(\mathcal{T}_q(\mathbf{r})\mathcal{R})]=\int_V \, d\mathbf{r} \, \oint \, d\mathcal{Q} \, \rho(\mathcal{Q}) \, [ \log \mathcal{V} \rho(\mathcal{Q}) - 1 \, ] \\ = V \, \oint \, d\mathcal{R} \, \rho(\mathcal{R}) \, [ \log \mathcal{V} \rho(\mathcal{R}) - 1 \, ] \, .
\end{multline}
We can therefore conclude that the ideal term is independent from $q$.

We now consider the excess free-energy term $\mathcal{F}_{ex}[\rho]$ and we describe an approach to minimize it exactly without recurring to the second-order $q$-expansion. The second-virial excess free-energy functional, with Parsons-Lee correction, for a chiral nematic phase reads
\begin{multline}
\beta \mathcal{F}_{ex}[\rho]= -\frac{G(\eta)}{2} \int_V \, d\mathbf{r} \, d\mathbf{r'}  \oint \, d\mathcal{R} \, d\mathcal{R}' \\
\times f(\mathbf{r}-\mathbf{r'},\mathcal{R},\mathcal{R'}) \, \rho(\mathcal{T}_q(\mathbf{r})\mathcal{R}) \, \rho(\mathcal{T}_q(\mathbf{r}')\mathcal{R}').
\end{multline}
By transforming the particle position variables and performing a volume integration, we can rewrite the previous equation as
\begin{multline}
\label{eq:chiralfex}
\frac{\beta \mathcal{F}_{ex}[\rho]}{V}= -\frac{G(\eta)}{2} \int_V \, d\mathbf{r} \oint \, d\mathcal{R} \, d\mathcal{R}' \\ 
\times f(\mathbf{r},\mathcal{R},\mathcal{R'}) \, \rho(\mathcal{T}_q(\mathbf{r})\mathcal{R}) \, \rho(\mathcal{R}').
\end{multline}
To extract the $q$-dependence from the density distribution, we expand $\rho$ in rotational invariants. For an \emph{achiral} nematic phase the expansion is
\begin{equation}
\rho(\mathcal{R})=\sum_{l=0}^{\infty} \sum_{m,n=-l}^{l} \rho^{l}_{mn} \mathcal{D}_{mn}^{l}(\mathcal{R}) \, ,
\end{equation}
where $\mathcal{D}_{mn}^{l}(\mathcal{R})$ are Wigner matrices~\cite{tung} and the expansion amplitudes read
\begin{equation}
\rho^{l}_{mn}=\frac{2l+1}{8\pi^2} \oint d\mathcal{R} \rho(\mathcal{R}) \mathcal{D}_{mn}^{l}(\mathcal{R})  \, .
\end{equation}
Similarly, for a chiral nematic phase we have
\begin{equation}
\label{eq:generalexp}
\rho(\mathcal{T}_q(\mathbf{r})\mathcal{R})=\sum_{l=0}^{\infty} \sum_{m,n=-l}^{l} \rho^{l}_{mn} \mathcal{D}_{mn}^{l}(\mathcal{T}_q(\mathbf{r})\mathcal{R}) \, .
\end{equation}
By inserting Eq.~(\ref{eq:generalexp}) into $\mathcal{F}_{ex}$ of Eq.~(\ref{eq:chiralfex}) we have
\begin{equation}
\label{eq:generalfex}
\frac{\beta \mathcal{F}_{ex}[\rho]}{V}= \frac{G(\eta)}{2} \, \sum_{l,m,n} \, \sum_{l',m',n'} \,  E_{mm'nn'}^{ll'} (q) \rho_{mn}^{l} \rho_{m'n'}^{l'}
\end{equation}
where we introduced the rotational-invariant $q$-dependent excluded-volume coefficients
\begin{multline}
\label{eq:generalexcl}
E_{mm'nn'}^{ll'} (q) =  - \int d\mathbf{r} \, \oint d\mathcal{R} \, d\mathcal{R'} \\
\times f(\mathbf{r},\mathcal{R},\mathcal{R}') \mathcal{D}_{mn}^{l}(\mathcal{T}_q(\mathbf{r})\mathcal{R}) \mathcal{D}_{m'n'}^{l'}(\mathcal{R}').
\end{multline}

Using Eq.~(\ref{eq:chiralfid}) and Eq.~(\ref{eq:generalfex}), we have thus shown that the free-energy functional of a chiral nematic phase can in general be written as
\begin{multline}
\label{eq:generalchdft}
\frac{\beta\mathcal{F}_q[\rho(\mathcal{R})]}{V}= \oint\,d\mathcal{R}\,\rho(\mathcal{R}) [ \log\rho(\mathcal{R})\mathcal{V} -1 ] \\
+ \frac{G(\eta)}{2} \sum_{l,m,n} \, \sum_{l',m',n'} E_{mm'nn'}^{ll'}(q) \rho_{mn}^l \rho_{m'n'}^{l'}.
\end{multline}
As a result, the $q$-dependence has been shifted from the ODF to the excluded volume coefficients.
Minimizing with respect to $\rho_{mn}^l$ and $q$ would allow to obtain the equilibrium properties of a general chiral nematic phase.

\subsection{Local uniaxiality}
\label{sec:localuni}

Let us now consider the simplest case of a chiral nematic phase: a phase which at $\mathbf{r}=0$ is locally uniaxial along the $z$ direction and is invariant to rotations around the main particle axis. In this case
\begin{equation}
\rho_{mn}^l = \sqrt{\frac{2l+1}{2}} \rho_l \delta_{m0} \delta_{n0} \, ,
\end{equation}
where the $l$-dependent factor is introduced for later convenience. The excess free energy of Eq.~(\ref{eq:chiralfex}) can therefore be expressed as
\begin{equation}
\frac{\beta \mathcal{F}_{ex}[\rho]}{V}= \frac{G(\eta)}{2} \sum_{l,l'=0}^{\infty}  \rho_l \rho_{l'} E_{ll'} (q) \, ,
\end{equation}
where $E_{ll'}(q)= \sqrt{\frac{2l+1}{2}} \sqrt{\frac{2l'+1}{2}} E_{0000}^{ll'} (q)$, which from Eq.~(\ref{eq:generalexcl}) reads
\begin{multline}
E_{ll'} (q) =  - \sqrt{\frac{2l+1}{2}} \sqrt{\frac{2l'+1}{2}} \int d\mathbf{r} \, \oint d\mathcal{R} \, d\mathcal{R'} \\
\times f(\mathbf{r},\mathcal{R},\mathcal{R}') \mathcal{D}_{00}^{l}(\mathcal{T}_q(\mathbf{r})\mathcal{R})\,  \mathcal{D}_{00}^{l'}(\mathcal{R}') .
\end{multline}
We can rewrite the rotational invariants in terms of the (standard) normalized Legendre polynomial $\mathcal{P}_{l}(x)$ (with $l$ the degree of the polynomial):
\begin{equation}
\mathcal{P}_{l}(\mathbf{\hat{n}} \cdot \mathbf{\hat{\omega}} ) = \sqrt{\frac{2l+1}{2}} \mathcal{D}_{00}^{l}(\mathcal{R}).
\end{equation}
It follows that
\begin{equation}
\rho_l = \int_{-1}^{1} \, d(\mathbf{\hat{n}} \cdot \mathbf{\hat{\omega}}) \,\rho(\mathbf{\hat{n}} \cdot \mathbf{\hat{\omega}}) \mathcal{P}_{l}(\mathbf{\hat{n}} \cdot \mathbf{\hat{\omega}}),
\end{equation}
which allows us to write, for a chiral nematic phase, 
\begin{equation}
\mathcal{P}_l(\mathbf{\hat{n}}_q(y) \cdot \mathbf{\hat{\omega}}) = \sqrt{\frac{2l+1}{2}} \mathcal{D}_{00}^l(\mathcal{T}_q(\mathbf{r})\mathcal{R}) .
\end{equation}
In the case of a cholesteric phase, in which the chiral director $\mathbf{\hat{\chi}} \perp \mathbf{\hat{n}}$ (and $\mathbf{\hat{\chi}} \parallel \mathbf{\hat{y}}$ in our reference frame), we have
\begin{equation}
\label{eq:chdirfield}
\mathbf{\hat{n}}_q (y) = \mathbf{\hat{x}} \, \sin qy \,   +  \mathbf{\hat{z}} \, \cos qy.
\end{equation}
Therefore, we obtain
\begin{multline}
\label{eq:exclvol}
E_{ll'} (q) =  - \int d\mathbf{r} \, \oint d\mathcal{R} \, d\mathcal{R'} \\
\times f(\mathbf{r},\mathcal{R},\mathcal{R}') \mathcal{P}_{l} (\mathbf{\hat{n}}_q (y) \cdot \mathbf{\hat{\omega}}) \mathcal{P}_{l'} (\mathbf{\hat{n}}_0 \cdot \mathbf{\hat{\omega}}')
\end{multline}
where $\mathbf{\hat{n}}_0 = \mathbf{\hat{n}}_q (0)$.
The excluded volume coefficients of Eq.~(\ref{eq:exclvol}) can be directly calculated using numerical techniques, as we explain in Sec.~\ref{sec:numerical}.

In conclusion, starting from Eq.~(\ref{eq:generalchdft}), valid for a generic chiral nematic phase, we have assumed local uniaxiality and independence of the distribution on rotations around the main particle axis, to obtain an explicit functional for the cholesteric phase, that we can rewrite as
\begin{multline}
\label{eq:cholfree}
\frac{\beta \mathcal{F}_q[\psi]}{V}= n (\log \mathcal{V}n -1)+ 4\pi^2 n \int_{-1}^{1} \, d\cos\theta \, \psi(\theta) \, \log \psi(\theta) \, \\
+ \frac{n^2G(\eta)}{2} \sum_{l,l'=0}^{\infty} \psi_{l} \psi_{l'} E_{ll'}(q),
\end{multline}
where, in analogy with the uniaxial case (cf. Eq.~(\ref{eq:onsager})), we kept only the dependence on the polar angle $\theta$ by defining the ODF $\psi(\theta)=\rho(\mathcal{R})/n$ and its expansion coefficients
\begin{equation}
\psi_{l}=\int_{-1}^1 \, d\cos\theta \, \psi(\cos\theta) \mathcal{P}_l (\cos\theta) \, .
\end{equation}
Once the excluded volume coefficients $E_{ll'}(q)$ defined by Eq.~(\ref{eq:exclvol}) are known, the equilibrium ODF is obtained, in complete analogy with Sec.~\ref{sec:onsager}, by solving the following equation
\begin{equation}
\label{eq:cheq}
\psi(\cos\theta)= \frac{1}{Z} \exp \left[ -n \, G(\eta) \sum_{l,l'=0}^{\infty} \frac{E_{ll'} (q)}{4\pi^2}  \mathcal{P}_l (\cos\theta) \psi_{l'} \right],
\end{equation}
with $Z$ the normalization constant. The minimization procedure is performed by using a very fine grid for the angles $\theta$ to obtain the solution $\psi(\theta)$, at fixed number density $n$ and fixed chiral wave vector $q$. For each density, the value of $q$ that corresponds to the minimum value of the free energy is the equilibrium one and the corresponding equilibrium pitch of the cholesteric phase reads $P=2\pi/q$.

\subsection{Limit of weak chirality: Straley's theory}
\label{sec:straley}
The expression obtained by Straley for the cholesteric pitch of weakly-chiral infinitely-long hard helices,~\cite{straley} can be easily assessed within our theoretical framework. Let us expand a general functional $\mathcal{F}[\rho(\mathcal{T}_q(y)\mathcal{R})]$ to second-order in $q$:
\begin{multline}
\frac{\mathcal{F}[\rho(\mathcal{T}_q(y)\mathcal{R})]}{V}=\\
\frac{\mathcal{F}[\rho(\mathcal{R})]}{V}+K_T[\rho(\mathcal{R})]q+\frac{1}{2}K_2[\rho(\mathcal{R})]q^2 + \mathcal{O}(q^3) , 
\end{multline}
where we introduced the constants
\begin{equation}
\label{eq:kt}
K_T[\rho(\mathcal{R})]= \left. \frac{1}{V}\frac{d\mathcal{F}_{ex}[\rho(\mathcal{T}_q(y)\mathcal{R})]}{dq} \right|_{q=0} \, ,
\end{equation}
and
\begin{equation}
\label{eq:k2}
K_2[\rho(\mathcal{R})]=\left. \frac{1}{V}\frac{d^2\mathcal{F}_{ex}[\rho(\mathcal{T}_q(y)\mathcal{R})]}{dq^2} \right|_{q=0} \, .
\end{equation}
$K_T$ is usually called the \emph{chiral strength} since it must differ from zero to have macroscopic chiral order. $K_2$ is the twist elastic constant.~\cite{straley,degennes} According to Straley's theory, the phase is homogeneous with respect to the internal angle, implying that the non-cylindrically-symmetric character of the particle is averaged out. The equilibrium cholesteric pitch is $P=2\pi / q = -2 \pi K_2/K_T$. Since in general $K_2>0$, the handedness of the cholesteric phase depends on the sign of $K_T$. The explicit expression for $K_T$ depends on the theoretical framework adopted, see for example Refs.~[44,58]
, and within the second-virial approximation it reads
\begin{multline}
\label{eq:kt_expl}
K_T = -\frac{n^2}{2}  \int d\mathbf{r} \, \oint d\mathbf{\hat{\omega}} \, d\mathbf{\hat{\omega}'}\\ \times f(\mathbf{r},\mathbf{\hat{\omega}},\mathbf{\hat{\omega}'}) \, y  \, \omega_x \, \psi(\mathbf{\hat{n}}_0 \cdot \mathbf{\hat{\omega}}) \, \dot{\psi}(\mathbf{\hat{n}}_0 \cdot\mathbf{\hat{\omega}'}),
\end{multline}
where $\omega_x=\mathbf{\hat{\omega}} \cdot \mathbf{\hat{x}}$ and $\dot{\psi}(x)=\partial \psi(x) / \partial x$ indicates the derivative with respect to the function argument. By using Eqs.~(\ref{eq:kt}) and~(\ref{eq:k2}), we will assess quantitatively Straley's approach for hard-helix systems. Moreover, we will compare our results with those presented in Ref.~[58]
, where a sophisticated implementation of Straley's approach has been used.

\subsection{Numerical procedure}
\label{sec:numerical}
As input for our theory, we need to evaluate the excluded-volume coefficients defined in Eq.~(\ref{eq:exclvol}). To perform these calculations, we  use a Monte Carlo (MC) integration scheme that has several advantages. First of all, it is a very general approach suitable for several particle models.~\cite{liu,ibarra,nguyen,demichele2012macro,demichele2012soft,movahed} Moreover, MC integration is a robust method for functions with discontinuities and for integration regions with complicated boundaries,~\cite{press} as in this study where we integrate the Mayer function of complex shaped particles. In general, it is also computationally more efficient than standard quadrature methods for the evaluation of high-dimensional integrals and it is intrinsically parallelizable since it consists of uncorrelated calculations. Furthermore, by determining the associated statistical errors (that decay as $\sim 1/\sqrt{n_{MC}}$, with $n_{MC}$ the number of MC steps), it is easy to control the accuracy of the calculations, as we will show below in Fig.~\ref{fig:numerics}{\bf(b)}. Finally, we point out that we use the simplest brute-force method for now; the implementation of more sophisticated schemes could be beneficial~\cite{schultz2010,schultz2014,benjamin,allen_private} but this is left for future studies.

\begin{figure}[h!t]
\begin{center}
\includegraphics[width=0.5\textwidth]{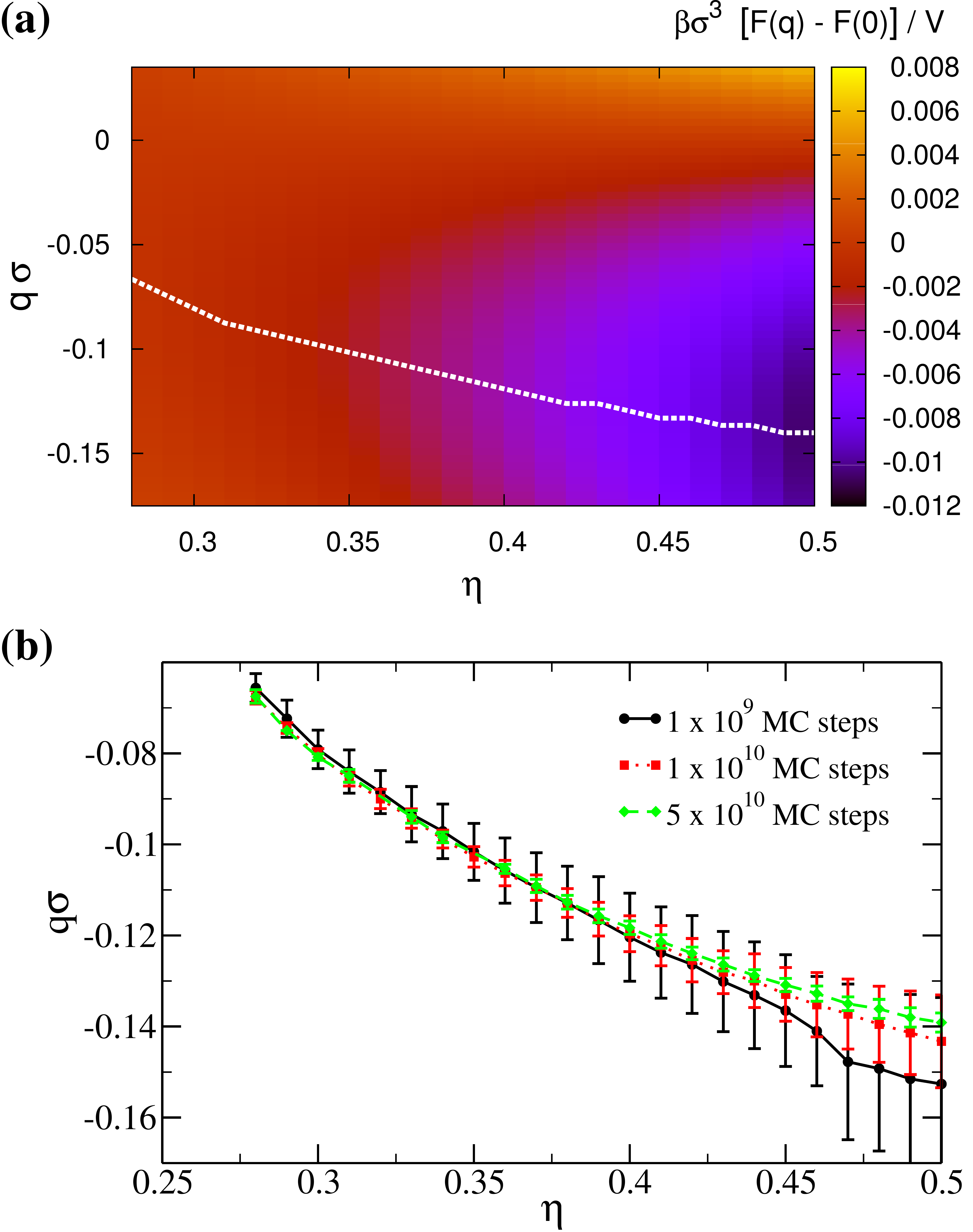}
\end{center}
\caption{Example of typical output of the theory, applied to a system of hard helices with molecular pitch $p=8\sigma$, outer radius $r=0.4\sigma$, contour length $L_c=10\sigma$, and number of beads (of diameter $\sigma$) $N_s=15$ (see Fig.~\ref{fig:micro_macro}{\bf(b)}). {\bf(a)} Free-energy difference $F(\eta,q)-F(\eta,0)$ between cholesteric and achiral nematic as a function of cholesteric wave vector $q$ and packing fraction $\eta$. The white dashed line indicates the free-energy minimum, thereby identifying the equilibrium $q$ for each $\eta$. {\bf(b)} Density-dependence of the cholesteric wave vector $q$ and associated statical errors computed over 8 independent runs of $n_{MC}=1\times10^9$ (black solid line), $n_{MC}=1\times10^{10}$ (red dotted) and $n_{MC}=5\times10^{10}$ (green dashed) MC steps. As expected, error bars are smaller for increasing number of integration steps $n_{MC}$. }
\label{fig:numerics}
\end{figure}

Assuming the first particle in the origin, the procedure consists of repeating $n_{MC}$ times the following steps: {\bf i)} generate uniformly the random variables $\mathbf{r}=(x,y,z) \in V$, $\theta , \theta' \in [0,\pi)$ and $\phi , \phi' , \alpha, \alpha' \in [0,2\pi) $ to obtain a random two-particle configuration; { \bf ii)} compute the Mayer function $f(\mathbf{r},\mathcal{R},\mathcal{R}')$, that in case of hard bodies consists of checking for particle overlaps; {\bf iii)} for each $l$, $l'$ and $q$ under consideration, compute the Legendre polynomials $\mathcal{P}_{l} (\mathbf{\hat{n}}_q (y) \cdot \mathbf{\hat{\omega}})$, $\mathcal{P}_{l'} (\cos\theta)$ and combine them with $f(\mathbf{r},\mathcal{R},\mathcal{R}')$ according to Eq.~(\ref{eq:exclvol}).
The coefficients $E_{ll'}(q)$ are therefore calculated as
\begin{multline}
E_{ll'}(q) = \\ - vol \langle f(\mathbf{r},\mathcal{R},\mathcal{R}') \mathcal{P}_{l} (\mathbf{\hat{n}}_q (y) \cdot \mathbf{\hat{\omega}}) \mathcal{P}_{l'} (\cos\theta) \sin\theta\sin \theta' \rangle_{MC}
\end{multline}
where for hard bodies the MC average $\langle . \rangle_{MC}$ reduces to an average over overlapping configurations, and $vol$ is the integration domain volume. For example, if positions are generated within a sphere of maximum radial distance $r_{max}$, we have $vol=(4/3)\pi r_{max}^3 (16\pi^6 )$. Coefficients up to $l_{max}=20$ are sufficient to ensure convergence, as also reported for spherocylinders.~\cite{cuetos} Calculations are sped up by using recursion formulas for $P_{l}$ and the number of coefficients $E_{ll'}$ can be reduced in case of additional symmetries. For example, up-down symmetric particles have only even coefficients $E_{2l2l'} \neq 0$.
After evaluating $E_{ll'}(q)$, we can iteratively solve Eq.~(\ref{eq:cheq}), at fixed number density $n$ and chiral wave vector $q$, by using a discrete grid for the polar angles $\theta$ (cf. Ref.~[69]
). The resulting ODFs are used to obtain a full free-energy landscape in the $(q,\eta)$ plane like the one shown in Fig.~\ref{fig:numerics}{\bf(a)}. Locating the free-energy minimum among the $q$-values studied at every packing fraction $\eta$, allows us to calculate the density-dependence of the equilibrium cholesteric pitch $P$. Pressure and chemical potential are derived from the free energy, and coexistence between isotropic and nematic phases is located by imposing equal pressure and equal chemical potential conditions. The most significant source of numerical errors in this procedure is caused by the limited accuracy due to a low number of $n_{MC}$ steps in evaluating the excluded-volume coefficients. However, by considering independent runs (that can subsequently be averaged to increase the accuracy) it is possible to carefully estimate the associated statistical errors. Fig.~\ref{fig:numerics}{\bf(b)}, for instance, shows the equilibrium cholesteric wave vector $q$ as a function of packing fraction $\eta$ for a system of hard helices, determined after the numerical calculation of the excluded volume coefficients with different number of MC steps $n_{MC}$. As a general trend, upon increasing the packing fraction $\eta$, errors become bigger since excluded-volume coefficients are coupled with density (see Eq.~(\ref{eq:onsager}) and Eq.~(\ref{eq:cheq})), with higher-$l$ coefficients (with poorer statistics) becoming increasingly important for the stronger peaked distributions at higher $\eta$. However, $n_{MC}$ can be increased in order to reach the desired accuracy. In the remaining, we will show the error bars only in a few cases, when we need to quantify our statistical accuracy before drawing conclusions on the physics of the system. The main drawback of the procedure is that a fine grid in $q$-values is computationally expensive and it is advisable to perform a shorter run in advance to define the optimal $q$-mesh.

\begin{figure*}[h!t]
\begin{center}
\includegraphics[width=\textwidth]{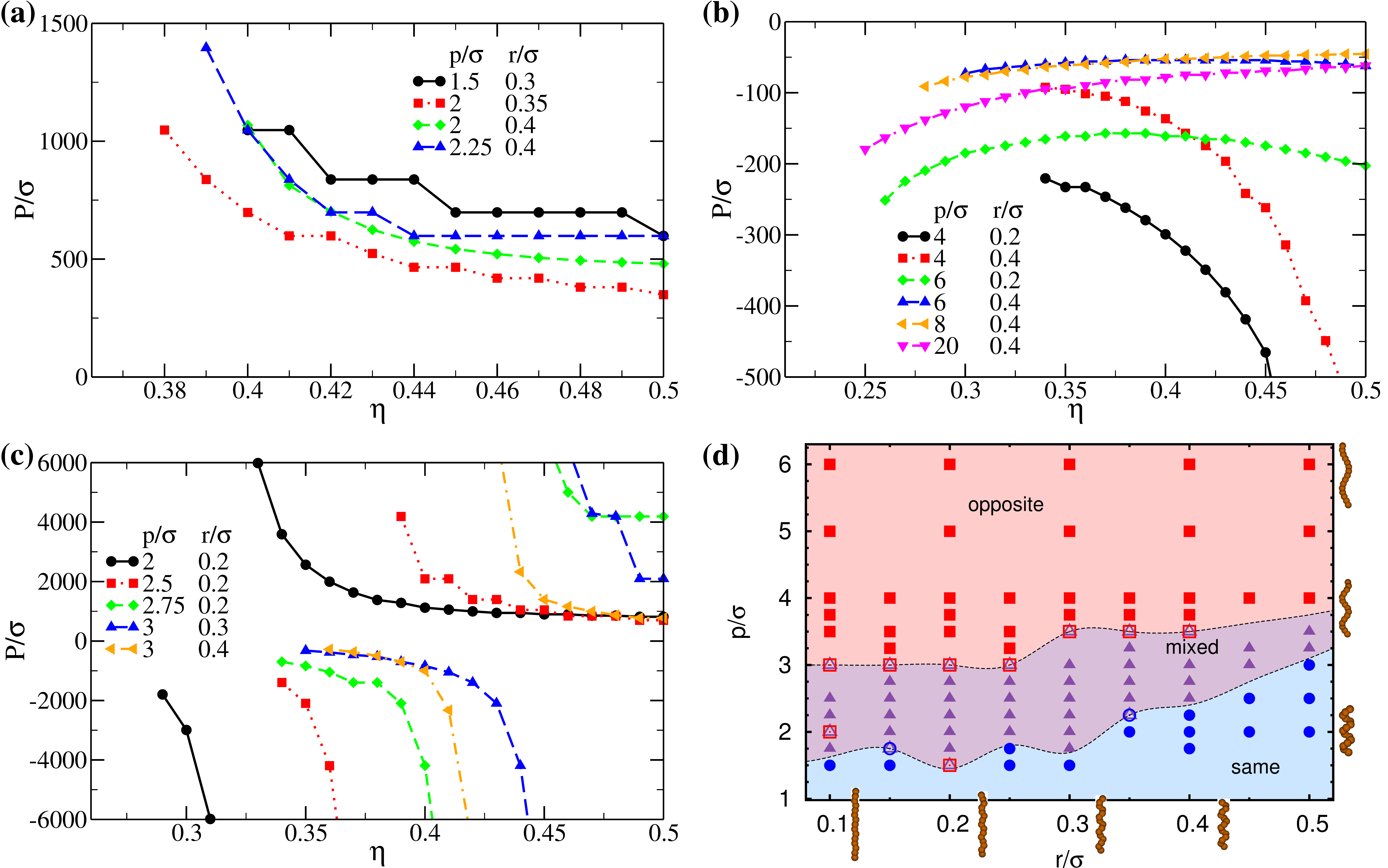}
\end{center}
\caption{Cholesteric pitch $P$ as a function of packing fraction $\eta$ for right-handed helices consisting of $N_s=15$ fused hard spheres with diameter $\sigma$, contour length $L_c=10\sigma$ and varying microscopic pitch $p$ and radius $r$ as labeled, stabilizing cholesteric phases with same {\bf(a)}, opposite {\bf(b)} or both {\bf(c)} handedness. {\bf(d)} State diagram for helices with $L_c=10\sigma$ and $N_s=15$ in the $p-r$ representation. Open symbols indicates parameters for which statistics is not sufficient for an accurate classification. Boundary lines are guides-to-the-eye.}
\label{fig:l10}
\end{figure*}

\begin{figure*}[h!t]
\begin{center}
\includegraphics[width=\textwidth]{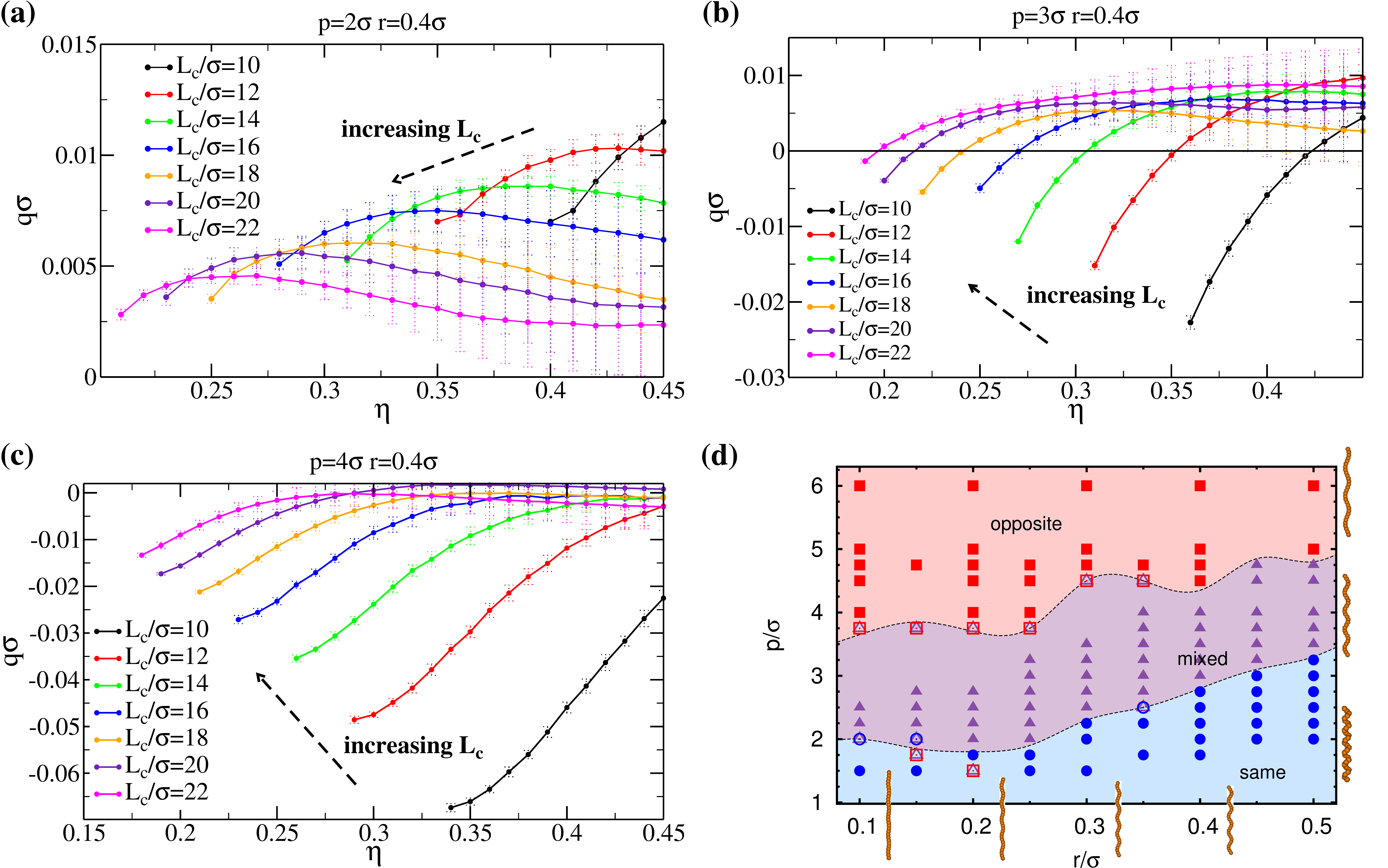}
\end{center}
\caption{Density dependence of the cholesteric wave vector $q$ for helices of fixed radius $r=0.4 \sigma $ and different contour lengths $L_c$ ($N_s=\frac{3}{2}L_c/\sigma$) with particle pitches {\bf(a)} $p=2\sigma$ {\bf(b)} $p=3\sigma$ {\bf(c)} $p=4\sigma$. {\bf(d)}  Radius $r$ - pitch $p$ state diagram for helices with contour length $L_c=20\sigma$ and $N_s=30$, with regions indicating the same, mixed, and opposite handedness regimes. Boundaries are shifted (upwards) with respect to the state diagram presented in Fig.~\ref{fig:l10}{\bf(d)}.}
\label{fig:length}
\end{figure*}

\begin{figure}[h!t]
\begin{center}
\includegraphics[width=0.5\textwidth]{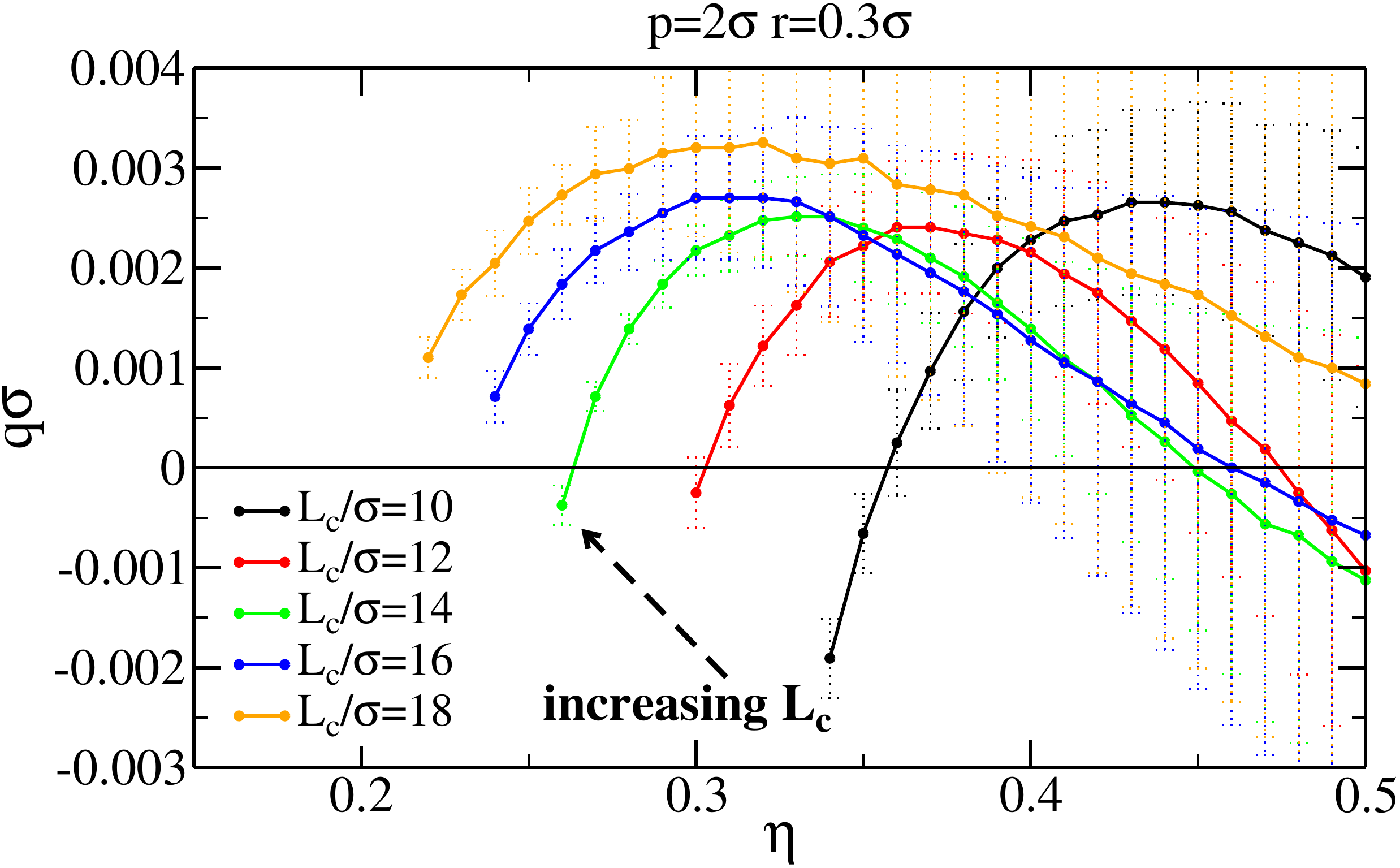}
\end{center}
\caption{Cholesteric wave vector $q$ as a function of packing fraction $\eta$ for helices with fixed geometry ($p=2\sigma$, $r=0.3\sigma$) and different contour length $L_c$ ($N_s=\frac{3}{2}L_c/\sigma$). Despite the large statistical error bars for large $\eta$, we will see in Fig.~\ref{fig:exclvol}{\bf(c)} that this trend is in principle possible.}
\label{fig:double}
\end{figure}

\section{Results}
\label{sec:results}
\subsection{Cholesterics of hard helices: handedness, (double) sense inversion, and length dependence}
\label{sec:hahe}

We study the cholesteric order arising in systems of colloidal hard helices. A hard helix is modeled as $N_s$ hard spheres of diameter $\sigma$, rigidly fused together to form an helix of contour length $L_c$, microscopic pitch $p$ and radius $r$ (see Fig.~\ref{fig:micro_macro}{\bf(b)} and Refs.~[49,63]
). In Fig.~\ref{fig:l10}{\bf(a)}-{\bf(c)}, we report the density dependence of the cholesteric pitch $P$ for right-handed helices with $L_c=10\sigma$ and $N_s=15$. We focus on the range of packing fractions for which the nematic phase is stable with respect to the isotropic phase, and we choose an upper limit of $\eta=0.5$ since at higher densities smectic phases are expected.~\cite{frezza2013,kolli2014jcp,kolli2014soft,bolhuis} Depending on the microscopic parameters $r$ and $p$, we observe three different cases. In Fig.~\ref{fig:l10}{\bf(a)}, we report the pitch $P$ of helices manifesting cholesteric phases with the same handedness of the constituent particles (positive pitch corresponds to right-handed twist). The magnitude of $P$ varies from hundreds to thousands of sphere diameters $\sigma$, depending on particle shape and system density, and it is monotonically decreasing upon increasing $\eta$. The step-like feature is an artifact due to the use of a discrete mesh for the $q$-values (see Sec.~\ref{sec:numerical}) and a smooth curve would be obtained by decreasing mesh-size and increasing statistics. Fig.~\ref{fig:l10}{\bf(b)} shows helices developing cholesterics with opposite handedness. The density dependence of $P$ appears to be more complex in this case but it is still possible to observe a few common trends. For instance, $|P|$ seems to exhibit a minimum for some $\eta$, and at fixed $p$, helices with a larger radius $r$ lead to a shorter cholesteric pitch $P$. It is worth noting that for some shapes, $|P|<100\sigma$, suggesting that these particle models would be good candidates for (direct) simulations of cholesteric phases (under twisted boundary conditions).
In Fig.~\ref{fig:l10}{\bf(c)}, we report cases in which the handedness of the cholesteric phases depends on the thermodynamic state of the system. For these peculiar helical shapes, a left-handed twist is preferred at the isotropic-cholesteric transition. However, upon increasing packing fraction, the cholesteric pitch $P$ becomes longer and longer, and passing via an achiral state (infinite $P$), eventually changes sense of twist. The packing fraction at which inversion occurs depends on the particle shape, and in general the inversion occurs at higher density for larger $p$ and $r$. Comparing these results with a recent study~\cite{wensink2014} that reported pitch inversion for long and soft (Yukawa) helices using Straley's approach, we observe that the nematic order parameter $S$ at which the inversion occurs can be much lower for the particles in the present study. Indeed, soft long helices~\cite{wensink2014} exhibit sense inversion only for a high degree of alignment ($S>0.9$), whereas for our short hard helices inversion can occur very close to the isotropic-cholesteric transition ($S\sim0.65$). If the corresponding packing fraction at which the inversion takes place becomes too high it is possible that another phase (e.g. smectic) becomes stable, thereby preventing such an inversion. To summarize our results, we report in Fig.~\ref{fig:l10}{\bf(d)} a state diagram using the molecular pitch $p$ and radius $r$ as axes of our representation. Depending on the functional behavior of $P$ vs $\eta$, we identify three regions that will be referred as \emph{same}, \emph{opposite} and \emph{mixed}. Open symbols represent cases for which our statistical accuracy is not enough for a precise classification. Uncertain points are also found for helices with very small $p$ and $r$ (e.g. $p=2\sigma$, $r=0.1\sigma$). For these parameters, the particle shape resembles a rod with protrusions rather than a proper helix, making the computation of the excluded volume more demanding and suggesting that very small changes in the shape give rise to complicated inter-locking effects.

In Fig.~\ref{fig:length} we study the dependence of the (inverse) pitch on particle contour length $L_c$ for selected particle shapes belonging to the three different classes (same, mixed, opposite). In Fig.~\ref{fig:length}{\bf(a)}, we report the cholesteric wave vector $q=2\pi/P$ for helices of fixed geometry ($p=2\sigma$ and $r=0.4\sigma$) and different length. We observe that an increase in $L_c$ corresponds to a decrease in $q$, therefore to a longer cholesteric pitch $P$ and a weaker cholesteric character. Upon increasing particle length, weaker cholesteric phases are also observed for helices undergoing handedness inversion (mixed case), as reported in Fig.~\ref{fig:length}{\bf(b)}. The same effect is also observed in helices stabilizing opposite cholesteric phases (see Fig.~\ref{fig:length}{\bf(c)}). We notice that $L_c$ does not only influence the magnitude of the cholesteric pitch but eventually also the sign and therefore the qualitative chiral behavior, even if the particle geometry is fixed (cf. also discussion on inclination angle in Ref.~[58]
). We summarize our results for helices of $L_c=20\sigma$ in the state diagram of Fig.~\ref{fig:length}{\bf(d)}, where we indeed observe an overall upward shift to higher values of $p$ of the boundaries delimiting the three different regimes with respect to Fig.~\ref{fig:l10}{\bf(d)}. Also in this case, the undulatory nature of the boundaries and the presence of unclear cases for small $p$ and $r$ reflect the sensitivity of the macroscopic chiral behavior on subtle changes in particle shape. We notice that neither the theoretical results obtained for long helices,~\cite{kornyshev2002,harris1997} predicting that $P \sim L^2$, nor experimental observations on coated fd viruses,~\cite{grelet2003} for which $P \sim L^{-0.25}$, are consistent with our study. Indeed, the richer scenario of short helices does not allow to deduce a clear scaling relation between cholesteric pitch $P$ and particle length $L$ (notice that $L\propto L_c$ for fixed $p,r$, see Eq.~(\ref{eq:LLc})).

To further emphasize that the chiral behavior of a system depends sensitively on the precise details of the single-particle properties, we conclude this section by speculating on a possible non-trivial cholesteric behavior as a function of particle length. In Fig.~\ref{fig:double}, we show the cholesteric wave vector $q$ as a function of packing fraction $\eta$, for helices with $p=2\sigma$, $r=0.3\sigma$ and various particle lengths. For $L_c=10\sigma$ we observe the (single) handedness inversion as described above. Surprisingly, upon increasing the particle length by a few $\sigma$, a second twist inversion seems to occur at higher packing fraction. However, we notice that large statistical errors are present at large $\eta$. Upon further increasing the particle length ($L_c=16\sigma$), the first inversion disappears. Therefore, in contrast with the previous case, the chirality inversion involves a transition from same to opposite handedness, upon increasing packing fraction. Finally, for helices with $L_c=18\sigma$ no inversion is present and only cholesterics with the same handedness are stable in the range of $\eta$ studied. Even though the large statistical uncertainty seems to undermine the conclusiveness of our observations, we will see in the next section that this behavior is consistent with our interpretation of the chiral order in terms of minimization of the excluded volume. However, it is not possible to exclude that another phase (e.g. smectic) would be more stable than the cholesteric at large $\eta$, preventing in particular the second inversion to occur.

\subsection{Competition between shape and particle-particle correlations}
\label{sec:interpretation}
\begin{figure}[h!t]
\begin{center}
\includegraphics[width=0.5\textwidth]{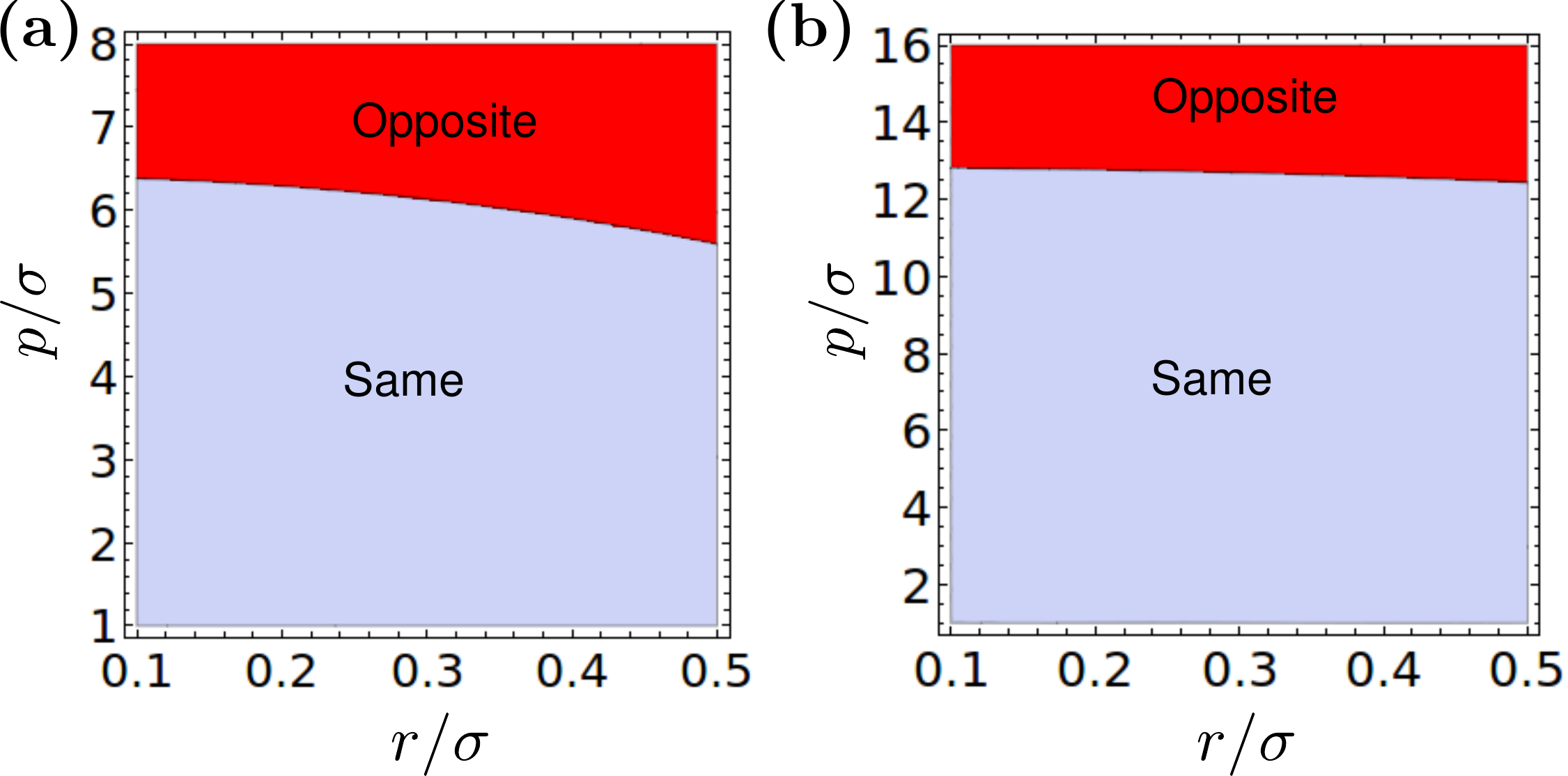}
\end{center}
\caption{State diagrams based on the sign of the pseudo-scalar $\psi_{HKL}$.~\cite{harris1997,harris1999} {\bf(a)} Hard helices with fixed contour length $L_c=10\sigma$ (cf. Fig.~\ref{fig:l10}{\bf(d)}). {\bf(b)} $L_c=20\sigma$ (cf. Fig.~\ref{fig:length}{\bf(d)}). Notice the different scales on the vertical axes.}
\label{fig:hkl}
\end{figure}

In this section we try to interpret our results on the collective chiral behavior in terms of a microscopic parameter. Harris, Kamien and Lubensky (HKL) proposed~\cite{harris1997,harris1999} a pseudoscalar $\psi_{HKL}$ to measure the internal chiral strength of a molecule, and showed that $\psi_{HKL}$ is proportional to the macroscopic chiral strength $K_T$, defined in Eq.~(\ref{eq:kt}). The sign of $\psi_{HKL}$ determines the handedness of the cholesteric phases. Ref.~[61]
 (see table I therein) reports an explicit expression of $\psi_{HKL}$ for an helix of uniform density, in the limit that the particle length is much larger than the particle radius. In our notation, it reads
\begin{equation}
\psi_{HKL} \propto -\frac{3r^4 L}{\left(2\pi \frac{L}{p}\right)^3} \left[ 1-\frac{24}{\left( \pi \frac{L}{p}\right)^2} \right] ,
\end{equation}
where $L$ is the Euclidean length, which is a function of the contour length $L_c$, the microscopic pitch $p$ and the radius $r$ given by
\begin{equation}
\label{eq:LLc}
L=\frac{p L_c}{2\pi \sqrt{r^2 +\left(\frac{p}{2\pi}\right)^2 } } .
\end{equation}
In Fig.~\ref{fig:hkl}, we report the state diagram for helices at fixed $L_c=10\sigma$ and $L_c=20\sigma$, based on the analysis of the sign of $\psi_{HKL}(r,p)$, in analogy with Figs.~\ref{fig:l10}{\bf(d)} and~\ref{fig:length}{\bf(d)}. Clearly, the mixed region where sense inversion occurs cannot be explained  by the pseudoscalar $\psi_{HKL}$, as it is density independent. Nevertheless, a qualitative trend can be captured with this purely geometric interpretation. We clearly find that the sign of $\psi_{HKL}$ overestimates significantly the value of the microscopic pitch $p$ for the boundary from same to opposite handedness, but describes correctly that this boundary shifts to higher $p$ upon increasing $L$. More over, the pseudoscalar approach predicts that the cholesteric pitch scales as $P \propto L^2$, which we have already shown not to be always the case for short helices. We can therefore conclude that single-particle properties are not sufficient to completely describe the observed non-trivial macroscopic chiral behavior and, as already noticed,~\cite{harris1997,harris1999} particle correlations must be taken into account as well.

\begin{figure*}[h!t]
\begin{center}
\includegraphics[width=\textwidth]{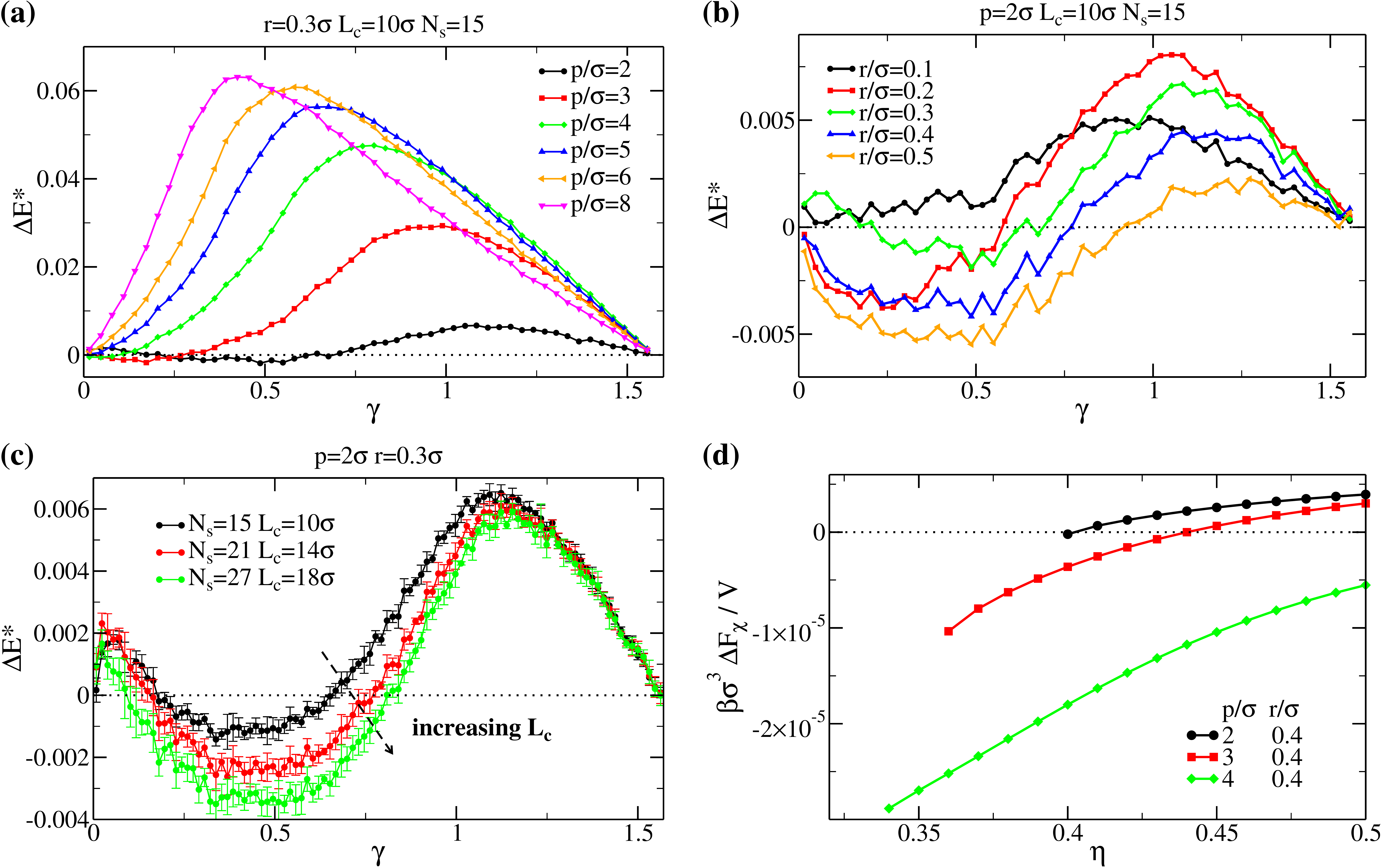}
\end{center}
\caption{Difference in excluded volume between right- and left-handed pair configurations $\Delta E^*=(E_R-E_L)/(E_R+E_L)$ as a function of the angle between the two helices $\gamma=\arccos(\mathbf{\hat{\omega}} \cdot \mathbf{\hat{\omega}'})$. {\bf(a)} Right-handed helices of length $L_c=10\sigma$, $N_s=15$, fixed radius $r=0.3\sigma$ and different microscopic pitch $p$. {\bf(b)} Right-handed helices of length $L_c=10\sigma$, fixed pitch $p=2\sigma$ and different radius $r$. {\bf(c)} Right-handed helices with fixed geometry $p=2\sigma$, $r=0.3\sigma$ and different length $L_c$. Error bars are calculated over 5 independent runs of $2\times 10^{10}$ MC steps. {\bf(d)} Thermodynamic average of the excluded volume difference $\Delta F_{\chi}$ as a function of $\eta$ for helices of length $L_c=10\sigma$ ($N_s=15$), $r=0.4\sigma$ and different $p$. The trends of $\Delta F_{\chi}$ match qualitatively with the density dependence of the cholesteric wave vector $q$ (cf. Fig.~\ref{fig:length}). }
\label{fig:exclvol}
\end{figure*}

As shown in our previous study,~\cite{belli} in order to explain the stability of chiral ordering we analyze the excluded volume associated to right/left-handed pairs of particles and the handedness of the resulting cholesteric phase. A pair of helices is in a right-handed configuration if $(\mathbf{r}-\mathbf{r'}) \cdot (\mathbf{\hat{\omega}} \times \mathbf{\hat{\omega}'}) >0$. Vice versa, if the latter is negative it is in a left-handed configuration. Therefore, we can define a right/left-handed excluded volume as
\begin{multline}
E_{\bfrac{R}{L}}(\mathbf{\hat{\omega}} \cdot \mathbf{\hat{\omega}'})=-\int d (\Delta \mathbf{r}) \int_0^{2 \pi} \frac{d \alpha}{2\pi} \frac{d \alpha'}{2\pi} \times \\
\times f(\Delta \mathbf{r}, \mathcal{R}, \mathcal{R}') \, \Theta ( \pm \Delta\mathbf{r} \cdot (\mathbf{\hat{\omega}} \times \mathbf{\hat{\omega}'}) ),
\end{multline}
with  $\Theta(x)$ the Heaviside step function and $\alpha$ the internal angle (cf. Sec.~\ref{sec:dft}). If $\Delta E \equiv E_R-E_L >0$ a left-handed configuration is preferred, and if $\Delta E<0$ a right-handed one. It is worth noting that $\Delta E$ is a microscopic property of a pair of helices, while $\psi_{HKL}$ is a single-particle property. For convenience, we define a normalized $\Delta E^*=(E_R-E_L)/(E_R+E_L)$. In Fig.~\ref{fig:exclvol}{\bf(a)}-{\bf(c)}, we report $\Delta E^*$ for several helical shapes (all right-handed), as a function of the angle formed by the main axis of the two helices $\gamma= \arccos(\mathbf{\hat{\omega}} \cdot \mathbf{\hat{\omega}'})$.  In the case that $\Delta E^*$ has the same sign for all values of the angle $\gamma$, we can predict undoubtedly the handedness of the cholesteric phase. For example, in Fig.~\ref{fig:exclvol}{\bf(a)} we report helices with fixed radius $r=0.3\sigma$ and length $L_c=10\sigma$ and different $p$ (moving on a vertical line in the state diagram of Fig.~\ref{fig:l10}{\bf(d)}). For large $p$, $\Delta E^* >0$ $\forall \gamma$, consistently with the stabilization of a left-handed cholesteric (opposite case). The magnitude of the cholesteric pitch $P$ is also qualitatively related to the magnitude of $|\Delta E^*|$. On the other hand, for helices with smaller $p$ we observe that $\Delta E^* < 0$ for small angles $\gamma$. In this case the handedness of the liquid-crystalline phase cannot be predicted a priori. In fact, for this range of parameters we have shown (cf. Fig.~\ref{fig:l10}{\bf(d)}) that the cholesteric handedness depends on packing fraction, i.e. it depends on the local alignment (mixed case). The pitch inversion can be qualitatively interpreted as follows. At low packing fraction, the average angle $\gamma$ between helices is relatively large and since the corresponding $\Delta E^*>0$ an opposite-handed phase is stabilized. Increasing the packing fraction, the average $\gamma$ becomes smaller and eventually $\Delta E^* < 0$, giving rise to a same-handed phase. The subtle balance between excluded volume and local alignment can also be appreciated in Fig.~\ref{fig:exclvol}{\bf(b)}, where we show results for helices with fixed internal pitch $p=2\sigma$ but different $r$ (horizontal line in Fig.~\ref{fig:l10}{\bf(d)}). Analyzing the state diagram, we find upon increasing $r$: an (uncertain) opposite case ($r=0.1\sigma$), mixed cases ($r/\sigma=0.2,0.3$), and same-handedness cases ($r/\sigma=0.4,0.5$). Observing that $\Delta E^*>0$ $\forall \gamma$ for helices with $r/\sigma=0.1$, we can confirm the opposite handedness in the state diagram. Such a behavior seems an anomaly in the state diagram but, as already mentioned, the helical shape in this region (small $p$ and small $r$) has complex features that can give rise to a non-ordinary behavior. We notice that also in the analysis of the maximum packing fraction performed in Ref.~[64]
 (cf. Fig. 3 therein), no clear trend can be observed for helices with small $p$ and small $r$ (e.g. helices with $p=1\sigma$ $r=0.2\sigma$ represent a local minimum in the maximum $\eta$). The angular dependence of $\Delta E^*$ for $r=0.2\sigma$ is already described above and explains the mixed case. A double inversion seems also possible in the case of $r=0.3\sigma$ and it will be described in detail in Fig.~\ref{fig:exclvol}{\bf(c)}. For helices with $r/\sigma=0.4,0.5$, we also observe two regions for $\Delta E^*$ but in these cases the range of angles $\gamma$ with $\Delta E^*<0$ is larger than the (mixed) case of $r/\sigma=0.2$, and $\Delta E^*>0$ is only for $\gamma$ values that are so large that they are not expected to contribute to the nematic order. Moreover, $\Delta E^*$ takes also values more negative. As a consequence, the stabilized phase has the same (right-handed) handedness of the constituent helices. The dependence of $\Delta E^*(\gamma)$ on particle length for $p=2\sigma$ and $r=0.3\sigma$, reported in Fig.~\ref{fig:exclvol}{\bf(c)}, is consistent with the observations made for the chiral behavior shown in Fig.~\ref{fig:double}. In all cases we observe for increasing $\gamma$ the sequence positive-negative-positive for $\Delta E^*$ suggesting the possibility of a double inversion for all particle lengths. However, the depth of the region in which $\Delta E^* <0$ is larger for longer helices, and is caused by the different chiral behavior for particles with different lengths, as already seen before. These observations manifest the intricate link between microscopic and macroscopic chirality. This indicates that although the calculation of $\Delta E^*$ is a powerful tool, and computationally faster than the full minimization, it cannot always be considered as an exhaustive analysis, which is to be expected as this type of analysis is based on geometrical two-body properties only, which do not take into account the thermodynamic state point. In order to do so, we introduce the ODF $\psi(\theta)$ to explicitly account for the local alignment in the system. We therefore \emph{thermodynamically} average the difference in the excluded volume, introducing a quantity that mimicks the functional form of the excess second-virial free energy:~\cite{belli}
\begin{multline}
\frac{\beta \Delta F_{\chi}}{V} =- \frac{n^2}{2} \oint d \mathbf{\hat{\omega}} \oint d \mathbf{\hat{\omega}'}\\
\times \, \psi_{0} (\mathbf{\hat{n}}_0 \cdot \mathbf{\hat{\omega}}) \psi_{0} (\mathbf{\hat{n}}_0 \cdot \mathbf{\hat{\omega}'}) \Delta E(\mathbf{\hat{\omega}}\cdot\mathbf{\hat{\omega}'}) ,
\end{multline}
with $\psi_{0} (\mathbf{\hat{n}}_0 \cdot \mathbf{\hat{\omega}'})$ the ODF in the achiral limit. In Fig.~\ref{fig:exclvol}{\bf(d)} we report $\Delta F_{\chi}$ for three representative cases of helices with $L_c=10\sigma$, $r=0.4\sigma$ and $p=2\sigma$ (same), $p=3\sigma$ (mixed) and $p=4\sigma$ (opposite). By comparing with the density dependence of $q$ in Fig.~\ref{fig:length} (black lines in panels {\bf (a)-(c)} respectively), we observe that all the three regimes are captured by $\Delta F_{\chi}$, including a good agreement on the packing fraction at which the sign inversion is obtained. We can therefore conclude that a solely geometric interpretation is not sufficient to describe our results and that the degree of local alignment must be taken into account by weighting the excluded volume difference $\Delta E^*$ with the ODF.

\subsection{Chiral order vs uniaxial order: weak chirality limit and comparison with Straley's approach }
\label{sec:comparisonstraley}
\begin{figure}[h!t]
\begin{center}
\includegraphics[width=0.5\textwidth]{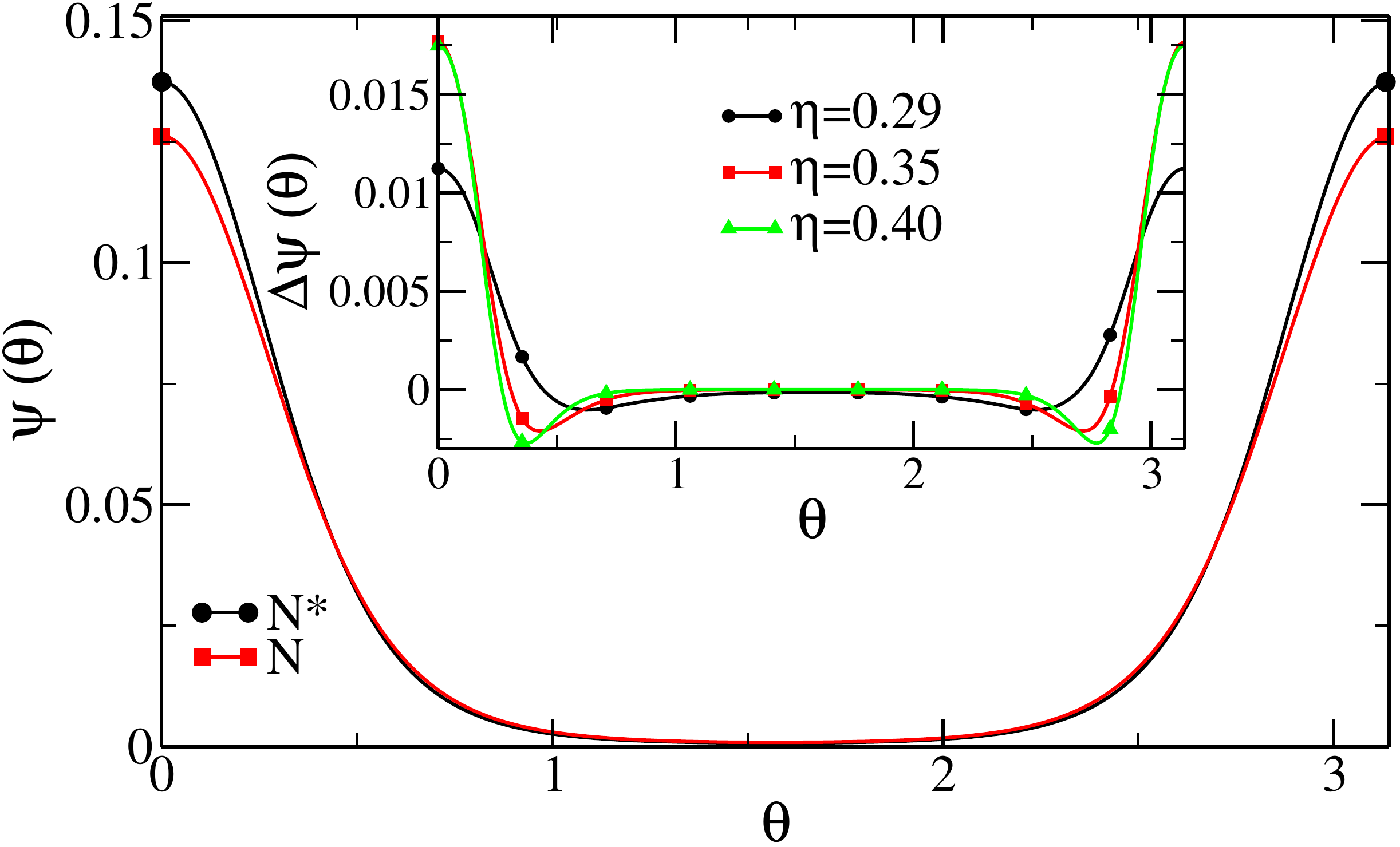}
\end{center}
\caption{Orientation distribution function (ODF) $\psi(\theta)$ for uniaxial nematic phase ($N$) and cholesteric ($N^*$) for helices with $p=8\sigma$, $r=0.4\sigma$, $L_c=10\sigma$, $N_{s}=15$ at the isotropic-nematic transition ($\eta \simeq 0.29$). Inset: difference $\Delta \psi(\theta)=\psi_{N^*}(\theta)-\psi_{N}(\theta)$ for different packing fraction.}
\label{fig:chiral_order}
\end{figure}

We have shown in Sec.~\ref{sec:straley} that we can recover the theory proposed by Straley~\cite{straley} by expanding the full free-energy functional for small $q$. An interesting difference between Straley's small-$q$ expansion with coefficients evaluated in the achiral limit ($q=0$) and the present study involves the effect of $q$ on the ODF which is taken into account here and ignored in Straley's approach. In Fig.~\ref{fig:chiral_order}, we show an example of the difference between the ODF corresponding to the achiral limit ($q=0$) and that at $q\neq0$ for which the free energy has actually a minimum (helices with $p=8\sigma$, $r=0.4\sigma$, $L_c=15\sigma$, $N_s=15$). We observe a more peaked ODF associated to the chiral order ($N^*$)  than the achiral one (N). In the inset, we see that this difference becomes more pronounced upon increasing the packing fraction since the liquid crystal phase becomes more chiral (smaller $\mathcal{P}$) for this kind of helices. Even though the difference can be small, it is yet reflected in the free energy (cf. Fig.~\ref{fig:numerics}{\bf(a)}) and in the nematic order parameter $S$ that can differ by a few percent.

\begin{figure}[h!t]
\begin{center}
\includegraphics[width=0.5\textwidth]{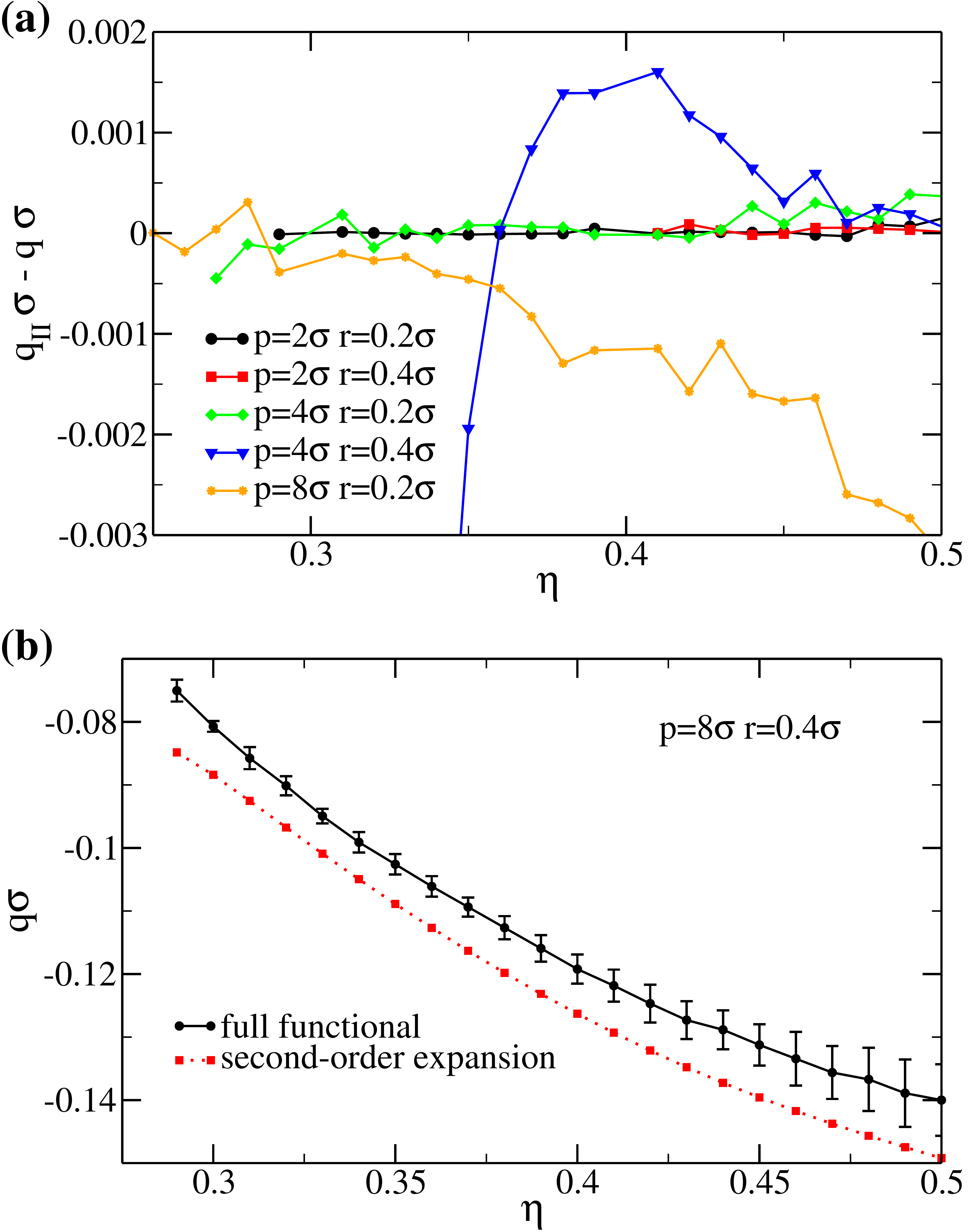}
\end{center}
\caption{{\bf(a)} Difference in the cholesteric wave vector $q_{II}-q$ as a function of $\eta$ obtained by expansion of the free-energy functional $q_{II}$ (see Eqs.~(\ref{eq:kt}) and~(\ref{eq:k2})) and by minimization of the full functional $q$, for selected helices of length $L_c=10\sigma$ and varying microscopic pitch $p$ and radius $r$. {\bf(b)} The cholesteric wave vector $q$ as a function of $\eta$ for helices with $p=8\sigma$, $r=0.4\sigma$, $L_c=10\sigma$.}
\label{fig:straley}
\end{figure}

\begin{table*}[h!t]
\begin{center}
\begin{tabular}{|c|c|c|c|c|c|c|c|c|c|c|}
\hline
{\bf helix} & $\eta_{N^*}$ & \multicolumn{2}{c|}{$S_{IN^*}$} & \multicolumn{2}{c|}{$10^4$ $K_T$} & \multicolumn{2}{c|}{$K_2$} & \multicolumn{2}{c|}{$P$} & $|\Delta P| / P$\\
\hline
& $\mbox{  }$ here $\mbox{  }$ & $\mbox{  }$ here $\mbox{  }$ & Ref.~[58] &  $\mbox{  }$ here $\mbox{  }$  & Ref.~[58] &  $\mbox{  }$ here $\mbox{  }$ & Ref.~[58] & $\mbox{  }$ here $\mbox{  }$  & Ref.~[58] & \\
\hline
$r=0.2$ $p=2$ & 0.300 & 0.699 &  0.64 & 4.27 & 4.83 & 0.199   & 0.154 & -2990 & -2008 & 30\%\\
\hline
$r=0.2$ $p=4$ & 0.274 & 0.677 & 0.66 & 47.3 & 41.36 & 0.194 & 0.177  & -260 & -268 & 3\% \\
\hline
$r=0.2$ $p=8$ & 0.258 & 0.690 & 0.68 & 42.3  & -29.03 & 0.203 & 0.184  & -310 & -399 & 28\%\\
\hline
$r=0.4$ $p=2$ & 0.403 & 0.612 & 0.60 & -10.7 & -3.83 & 0.160 & 0.153 & 965 & 2509 & 160\% \\
\hline
$r=0.4$ $p=4$ & 0.340 & 0.622 & 0.61 & 110  & 98.35 & 0.150 & 0.152 & -93 & -97 & 4\%\\
\hline
$r=0.4$ $p=8$ & 0.282 & 0.619 & 0.61 & 115 & 110.13 & 0.136 & 0.159 & -90 & -90 & 0\%\\
\hline
\end{tabular}
\end{center}
\caption{Comparison between the method described here and Straley's approach as implemented in Ref.~[58]
. Our results are obtained by averaging 16 runs of $10^{10}$ MC steps for the excluded volume integration (only significant digits are reported). $K_T$ and $K_2$ are calculated using Eqs.~(\ref{eq:kt}) and~(\ref{eq:k2}), and the cholesteric pitch $P$ by minimizing the full functional. All the quantities are in reduced units with $k_BT=1$ and $\sigma=1$.}
\label{tab:straley}
\end{table*}

In Straley's method~\cite{straley} the uniaxial ODF is then used to compute the chiral strength $K_T$ and the twist elastic constant $K_2$. Subsequently, the equilibrium cholesteric wave vector in the second-order expansion approximation is obtained via $q_{II}=-K_T/K_2$. In our case, from the calculated free energy landscape, by using Eqs.~(\ref{eq:kt}) and~(\ref{eq:k2}), we are able to obtain the density dependence of these two constants. In Fig.~\ref{fig:straley} we assess quantitatively Straley's method by plotting $q_{II}$ obtained from second-order expansion and $q$ obtained by the minimization of the full functional, for selected helical shapes (the ones studied in Ref.~[58]
). We observe that for the helices in Fig.~\ref{fig:straley}{\bf(a)} the difference is  very small. Since the macroscopic chiral behavior for these particles is very weak (cf. Sec.~\ref{sec:hahe}) we expected that a second-order approximation would not be too off. However, in case of cholesterics with shorter pitch we find an appreciable difference, as can be observed from Fig.~\ref{fig:straley}{\bf(b)} in which we report the case for helices with $p=8\sigma$, $r=0.4\sigma$. Since the higher-order terms can be positive or negative (without any clear correlation with the chiral behavior), we can not conclude that Straley's method under/overestimate the results systematically. In Tab.~\ref{tab:straley}, we compare our results with Ref.~[58]
, in which a sophisticated implementation of Straley's method was used. Surprisingly, we observe that for very long cholesteric pitch $P$, where we expected the second-order approximation to be more accurate, the values differ the most. This is probably due to the fact that $K_T$ is very small and therefore a large precision in estimating such a quantity is needed to prevent a huge discrepancy in the cholesteric pitch. We can thus confirm that the overall scenario seems to be well captured by Straley's approach. However, we can not exclude that subtleties in the numerical implementation could lead to quantitative discrepancy for the value of the cholesteric pitch $P$ in some particular cases (for example, cf. helices with $p/\sigma=2$ in Tab.~\ref{tab:straley}).

\begin{figure}[h!t]
\begin{center}
\includegraphics[width=0.5\textwidth]{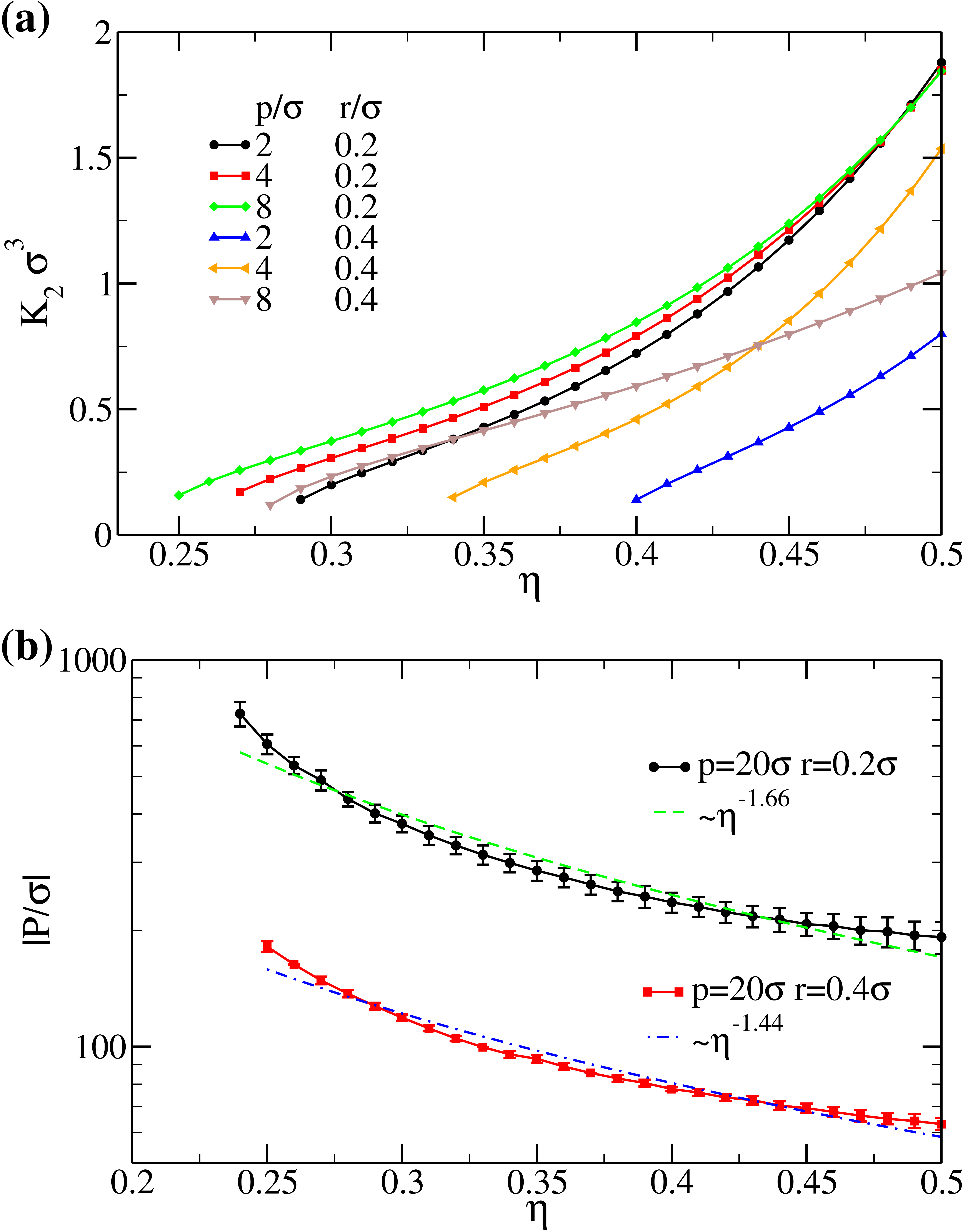}
\end{center}
\caption{{\bf(a)} Density dependence of the twist elastic constant $K_2$, calculated by using Eq.~(\ref{eq:k2}), for selected helices of length $L_c=10\sigma$. {\bf(b)} Log-Lin plot of absolute value of the cholesteric pitch $P$ for helices of length $L_c=10\sigma$ and very long internal pitch $p=20\sigma$. Lines are best fits of the power laws indicated in the legend.}
\label{fig:scaling}
\end{figure}

We conclude our analysis, by noting that in the limit of weakly chiral long helices, several studies predicted various scaling relations for the main quantities regulating the chiral liquid-crystalline behavior. Despite the relatively small $\eta$-regime of interest here, much smaller than a decade, which limits the meaning of exponents, we briefly discuss these scalings anyway for comparison with our results. For example, in his original work~\cite{straley} Straley proposed that $K_2 \sim \eta^2$, resulting in a cholesteric pitch $P \propto 1/S^2$. However, experiments often show a different density dependence of the elastic constant (for example in the thermotropic PBLG,~\cite{dupre} $K_2 \sim \eta^{0.36}$). Subsequently, Odijk~\cite{odijk} suggested that $P \sim \eta^{-1}$ for rigid hard helices, and $P \sim \eta^{-5/3}$ for flexible hard helices. Scaling relations are also measured in experiments, for example some fd viruses~\cite{grelet2003} show $P \sim \eta^{-1.45}$, while $P \sim \eta^{-1.8}$ is observed for PBLG.~\cite{dupre} In contrast, we find that none of these relations apply uniquely to the short hard helices studied here. Indeed, in Fig.~\ref{fig:scaling}{\bf(a)}, where we plot $K_2$ as a function of $\eta$ for several helical shapes, we observe that the functional form of the twist elastic constant $K_2$ computed from Eq.~(\ref{eq:k2}), depends on the different helical shape considered and cannot be described by a simple power law relation, at least not for the present parameter set. Analogously, in Fig.~\ref{fig:scaling}{\bf(b)}, we plot $P$ vs $\eta$ for very weakly chiral helices ($p=20\sigma$) along with the best fits of the expected power laws in these regimes. Due to the poor mutual agreement of the exponents, we tend to conclude that also the density dependence of the cholesteric pitch $P$ of short helices does not obey any general power law.

\begin{figure*}[h!t]
\begin{center}
\includegraphics[width=\textwidth]{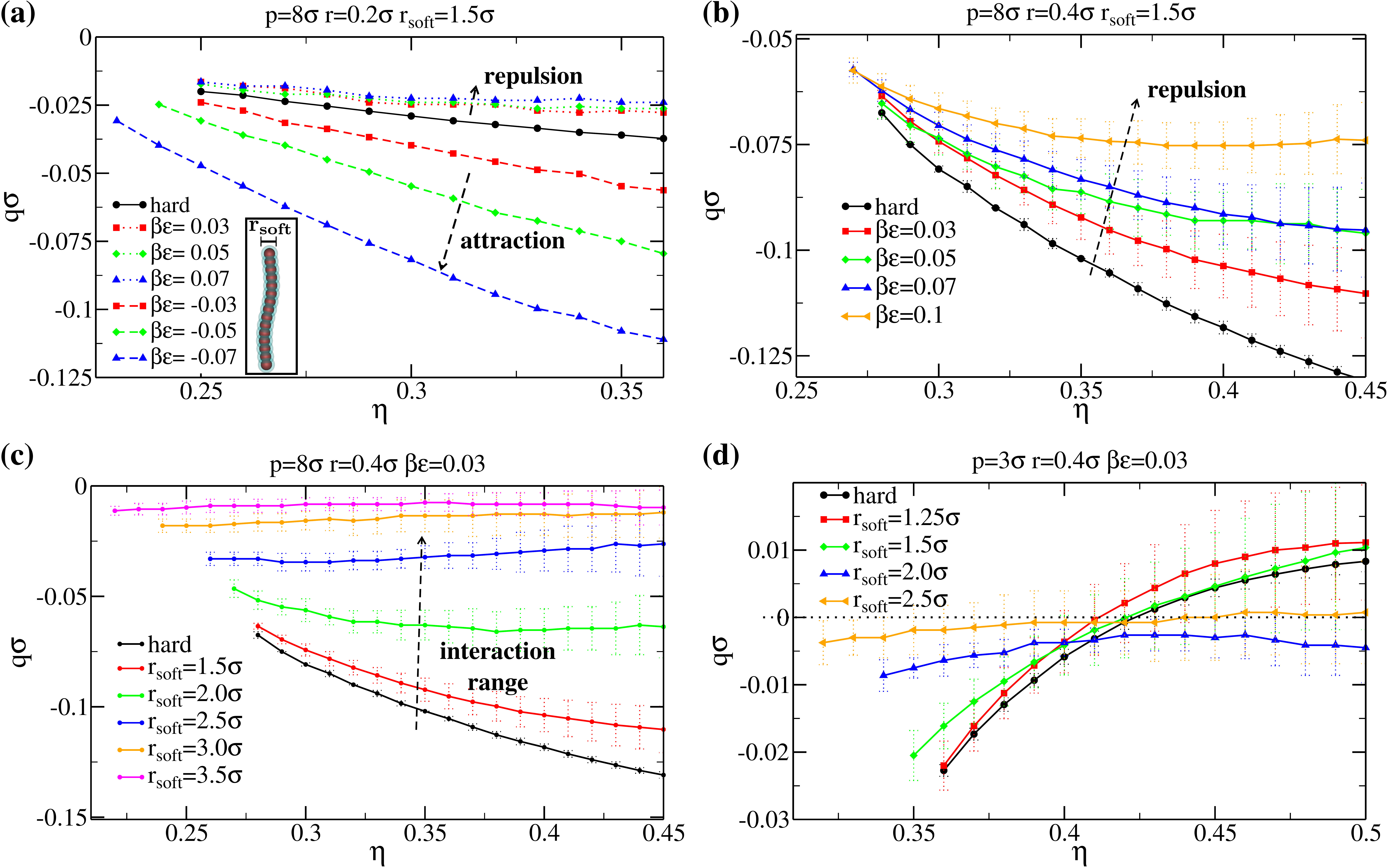}
\end{center}
\caption{Density dependence of the cholesteric wave vector $q$ for selected cases of short-ranged soft helices of length $L_c=10\sigma$. {\bf(a)} Effect of attraction ($\beta\epsilon<0$, dashed lines) and repulsion ($\beta\epsilon>0$, dotted lines) with respect to the hard case (full line) of helices with a microscopic pitch $p=8\sigma$, radius $r=0.2\sigma$ and interaction range $r_{soft}=1.5\sigma$. Inset: cartoon of the particle model. {\bf(b)} Effect of interaction strength $\beta\epsilon$ for short-range ($r_{soft}=1.5\sigma$) repulsive helices. {\bf(c)} Effect of interaction range $r_{soft}$ for helices with $p=8\sigma$, $r=0.4\sigma$ and interaction strength $\beta\epsilon=0.03$ stabilizing an opposite-handed cholesteric. {\bf(d)} Effect of interaction range $r_{soft}$ on helices ($p=3\sigma$, $r=0.4\sigma$, $\beta\epsilon=0.03$) exhibiting handedness inversion.}
\label{fig:soft}
\end{figure*}

\section{Towards more realistic particle models: softer colloids}
\label{sec:soft}

In this section, we modify the particle model to study the effect of an additional soft interaction on the macroscopic chiral behavior. The helical shape is still described by the parameters $N_s$, $L_c$, $p$ and $r$ (see Sec.~\ref{sec:hahe}), but spheres of different helices now attract or repel each other via the following short-range potential (cf. cartoon in Fig.~\ref{fig:soft}{\bf(a)}):
\begin{equation}
\beta U (r_{1i2j})= \left\{ 
\begin{array}{ll} 
\infty & \,\, r_{1i2j} \leq \sigma \\ 
\beta \epsilon & \,\, \sigma<r_{1i2j}<r_{soft} \\ 
0 & \,\, r_{1i2j} \geq r_{soft} 
\end{array} \right.
\end{equation}
where $r_{1i2j}$ is the distance between sphere $i$ of helix 1 and sphere $j$ of helix 2, and $r_{soft}$ determines the range of the potential. For $\beta\epsilon>0$ $(<0)$ we obtain a repulsive square shoulder (attractive square well) potential, whereas for $\beta\epsilon = 0$ we recover the hard-core potential studied above. Even though the computation of the excluded volume coefficients becomes more expensive, we are still able to obtain reliable results (cf. error bars in Fig.~\ref{fig:soft}) using the simple procedure described in Sec.~\ref{sec:numerical}.

In Fig.~\ref{fig:soft}{\bf(a)}, we report the density dependence of $q$ for helices with $p=8\sigma$, $r=0.2\sigma$, which interact via a very short-ranged potential ($r_{soft}=1.5\sigma$) that can be either attractive or repulsive. We observe that the effect of an additional attraction (repulsion) enhances (reduces) the macroscopic chiral behavior with respect to the purely hard helices. In fact, upon increasing the attractive well from $\beta\epsilon=0$ (hard case) to $|\beta\epsilon|=0.07$, the cholesteric pitch can be decreased by hundreds of $\sigma$, depending on $\eta$ as well. The opposite effect is obtained when the soft interaction is repulsive, as can be also observed in Fig.~\ref{fig:soft}{\bf(b)}. In this case ($p=8\sigma$, $r=0.4\sigma$), the soft repulsion with a range of $r_{soft}=1.5\sigma$ masks partially the molecular chiral features producing an effective shape that resembles more achiral rods. In Fig.~\ref{fig:soft}{\bf(c)}, we study at fixed interaction strength $\beta \epsilon=0.03$ the effect of the interaction range of the repulsion $r_{soft}$ for the same helices. Analogously, increasing the interaction range produces a longer cholesteric pitch, whose equilibrium value depends less sensitively on the density. Such an effect is also observed for helices manifesting both right- and left-handed phases (mixed case), as reported in Fig.~\ref{fig:soft}{\bf(d)}. In this particular case ($p=3\sigma$, $r=0.4\sigma$), the transition between the two types of cholesterics becomes less abrupt for increasing interaction range, eventually making it hard to identify within our statistical accuracy.

\begin{figure*}[h!t]
\begin{center}
\includegraphics[width=\textwidth]{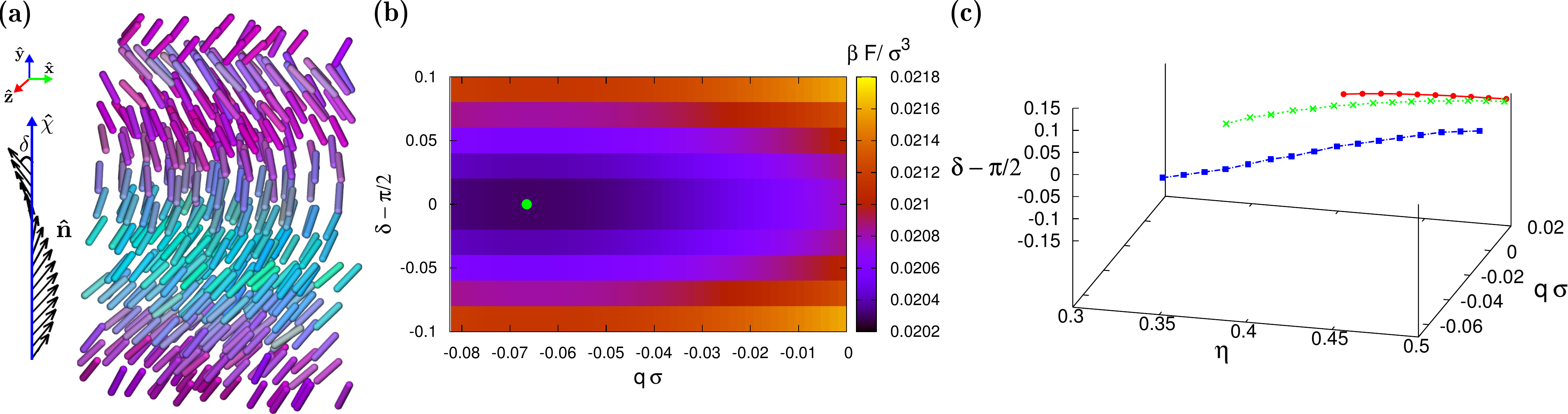}
\end{center}
\caption{{\bf(a)} Schematic of a twist-bend nematic phase. The local nematic director $\mathbf{\hat{n}}$ twists around the chiral director $\mathbf{\hat{\chi}}$ with a fixed bending angle $\delta$. {\bf(b)} Free-energy landscape in $\delta-q$ plane, at fixed packing fraction $\eta=0.35$ for helices with $p=4\sigma$, $r=0.4\sigma$ and $L_c=10\sigma$. The green circle indicates the free-energy minimum. {\bf(c)} Minima of free energy in $\eta-q-\delta$ space for helices of length $L_c=10\sigma$, $r=0.4\sigma$ and $p=2\sigma$ (red circles), $p=3\sigma$ (green crosses), $p=4\sigma$ (blue squares). All the lines belong to the plane $\delta=\pi/2$, indicating that bending is not favorable.}
\label{fig:tb}
\end{figure*}

\section{Towards more complex chiral phases: twist-bend nematics}
\label{sec:twistbend}
Cholesteric order is the simplest chiral arrangement for a nematic phase, in which the twist axis is perpendicular to the local nematic director.
In general, the local nematic director can be inclined by an angle $\delta$ with respect to the chiral director (see Fig.~\ref{fig:tb}{\bf(a)}), leading to the following functional dependence of the director field:
\begin{equation}
\label{eq:tbdirfield}
\mathbf{\hat{n}}_q (y) = \mathbf{\hat{x}} \, \sin\delta \sin qy + \mathbf{\hat{y}} \, \cos\delta + \mathbf{\hat{z}} \, \sin\delta \cos qy
\end{equation}
In Eq.~(\ref{eq:tbdirfield}) we keep on assuming the $y$-axis to be the chiral director. When $0 < \delta < \pi/2$, the phase is named twist-bend, or conical. This phase could even display additional hexatic order~\cite{kamien} that will not be taken into account here. The case of the simpler cholesteric is recovered for $\delta=\pi/2$, whereas for $\delta=0$ we have an achiral uniaxial nematic. By inserting Eq.~(\ref{eq:tbdirfield}) (instead of Eq.~(\ref{eq:chdirfield})) into the expression for the excluded volume coefficients (Eq.~(\ref{eq:exclvol})), our theory can be used to discriminate between cholesteric and conical phases. By implementing a 2D grid for the chiral wave vector $q$ and the angle $\delta$, we compute the excluded volume coefficients and calculate the free-energy landscape, that now depends on an additional parameter, the angle $\delta$ (cf. Fig.~\ref{fig:tb}{\bf(b)}).
In analogy with the cholesteric case, we are able to locate the free-energy minima in the $\eta-q-\delta$ space and obtain the equilibrium chiral properties characterizing the nematic phase. In Fig.~\ref{fig:tb}{\bf(c)}, we report the results for helices of contour length $L_c=10\sigma$, $r=0.4\sigma$ and $p/\sigma=2,3,4$. We recover the three regimes for the cholesteric handedness (same, mixed, opposite) and we find that the angle $\delta=\pi/2$ $\forall \eta$, indicating that bending in the system is not favorable. Therefore, we conclude that for these helical shapes the cholesteric phase is stable with respect to a twist-bend, or conical phase. Nevertheless, a thorough analysis should be performed, since evidences of a conical phase are reported in experiments on colloidal helical flagella,~\cite{barry2009} but none about a stable cholesteric phase, that could be hidden in the large isotropic-conical phase coexistence region. Clearly, the simple model presented in this paper might not be suitable to describe the experimental conditions of Ref.~[17]
. In particular, we notice that quite a large degree of polydispersity is observed, thus suggesting that deviations from the present monodisperse analysis might be found. Moreover, a screw-like nematic phase has been observed in simulations of hard helices,~\cite{kolli2014jcp} that differs from the conical one since the chiral arrangement refers to the short axis of the particles. However, due to the periodic boundary conditions used in the simulations, the stable cholesteric phase was not obtained, and therefore the competition between cholesteric and screw-like phases could not be examined. Indeed, the screw-like order could be seen as a partial manifestation of a biaxial chiral nematic phase. It will be intriguing to explicitly take into account the biaxial order in the second-virial theory described in this paper. Finally, the study of twist-bend order can be useful to give new insights into the intricate mechanisms governing liquid crystals formed by bent-core mesogens.~\cite{eremin}

\section{Concluding remarks}
\label{sec:conclusions}
We have developed a second-virial density functional theory for the chiral order in nematic phases. We set up a theoretical framework to obtain the equilibrium cholesteric pitch, eliminating (some of) the assumptions of Straley's approach.~\cite{straley} The use of MC integration as numerical method for the calculation of the effective excluded volume renders the theory fast, easy to implement and suitable for a wide range of particle models. We apply our theory to study the cholesterics of short hard helices, an apparently simple colloidal model that displays a richer chiral behavior than expected when considering long weakly chiral helices. In particular, we focus on the handedness of the cholesteric phase and we find a non-trivial dependence on particle shape and length, leading to a possible double sense inversion in some cases. We interpret our results as a competition between the geometric properties and the tendency of local alignment, resulting in a thermodynamic average of the difference in the excluded volume associated to right- and left-handed pairs. We also provide a quantitative comparison with Straley's theory, confirming that the most important features of the macroscopic chiral behavior can be captured with that method as well.
Our results provide new insights on the role of entropy in the link between micro- and macro-chirality, suggesting that entropy should not be overlooked in experiments on colloidal liquid crystals since most of the unexpected chiral phenomena could be ascribed to entropic effects only.
However, the limited aspect ratio of our particles and the lack of other important features, for example flexibility, have to be considered to fully analyse the phase behavior of some fd viruses.~\cite{dennison2011jcp,dennison2011prl,movahed,dogic2004,fynewever} By incorporating short-range soft interactions into the hard helix model, we have shown that it is possible to assess the macroscopic chiral behavior also beyond non purely hard-core colloids. However, it is likely that in order to deal with more complex inter-particle potentials, a more sophisticated implementation of MC integration should be considered.

The theoretical description of the chiral phase can also be improved. Since the biaxial order is expected to be strongly coupled to the chiral order,~\cite{harris1997,harris1999,priest,dhakal} future studies based on an orientation distribution function that explicitly accounts for the local biaxial arrangement, would provide new insights into the problem. Nevertheless, we have already shown that more complex chiral nematic phases, such as twist-bend nematic, sometimes called conical phase, can be straightforwardly studied within our theoretical framework. Additionally, introducing local biaxiality would allow us to better understand the competition in systems of hard helices between the cholesteric phase and the recently discovered screw-like phase.~\cite{kolli2014jcp,kolli2014soft} Furthermore, our theory can easily be extended to mixtures, addressing other fundamental questions such as the doping of achiral nematic phases and the chiral behavior of racemic mixtures, topics on which we are currently working. Finally, the recent progresses in chemical synthesis and in controlling the colloidal self-assembly processes, resulting in chiral superstructures, suggest that the number and variety of chiral building blocks will be soon enlarged.~\cite{zerrouki,yin,jiang,singh,smallenburg2012,pickett,chen,okazaki} Our approach will be useful to describe the macroscopic chiral behavior of these new colloids.

\section*{Acknowledgments}
We acknowledge financial support from a NWO-ECHO grant. This work is part of the research program of FOM, which is financially supported by NWO, and is part of the D-ITP consortium, a program of NWO funded by the Dutch Ministry of Education, Culture and Science (OCW). We thank SURFsara (www.surfsara.nl) for the support in using the Lisa Compute Cluster. We thank Mike Allen for useful correspondence and discussions.

\section*{References}
\bibliography{dussi_biblio}

\end{document}